\documentclass[aps,prl,showpacs,superscriptaddress,groupedaddress,nofootinbib]{revtex4-2}  
\usepackage{cancel}
\usepackage{graphicx} 
\usepackage[T1]{fontenc}
\usepackage{lmodern} 
\usepackage{graphicx}  
\usepackage{dcolumn}   
\usepackage{bm}        
\usepackage{amssymb}   
\usepackage{amsmath}
\usepackage{bbold}
\usepackage{array}
\usepackage{float}
\usepackage{makecell}
\usepackage{braket}
\usepackage{longtable}
\usepackage{supertabular,booktabs}
\usepackage{soul}

\usepackage{titlesec} 
\usepackage[colorlinks,citecolor=blue,urlcolor=blue,hypertexnames=true]{hyperref}
\usepackage{subfigure}

\usepackage[usenames,dvipsnames,svgnames,table]{xcolor} 

\renewcommand{\arraystretch}{2}

\newcommand{\be}{\begin{equation}}
\newcommand{\ee}{\end{equation}}
\newcommand{\bea}{\begin{eqnarray}}
\newcommand{\eea}{\end{eqnarray}}
\newcommand{\bml}{\begin{subequations}}
\newcommand{\eml}{\end{subequations}}
\newcommand{\bfig}{\begin{figure}}
\newcommand{\efig}{\end{figure}}

\newcommand{\bmat}{\begin{pmatrix}}
\newcommand{\emat}{\end{pmatrix}}
\usepackage{graphicx, slashed,booktabs, color, multirow, float,
amsfonts, bbold, mathtools, sidecap, tikz, bm,enumitem}
\usepackage{multirow}
\usepackage{bbding}
\usepackage{titlesec}
\usepackage{hyperref}
\usepackage{wasysym}
\usepackage{amssymb}
\usepackage{pifont}

\renewcommand{\leq}{\leqslant}

\usepackage[dvipsnames, usenames]{xcolor}

\definecolor{linkcolor}{rgb}{0.55, 0.13, .32}

\definecolor{oucrimsonred}{rgb}{0.6, 0.0, 0.0}
\definecolor{persianblue}{rgb}{0.11, 0.22, 0.73}
\definecolor{forestgreen}{rgb}{0.13,0.35,0.13}
\definecolor{lightgray}{rgb}{0.83, 0.83, 0.83}
 \hypersetup{colorlinks, citecolor=oucrimsonred, linkcolor=persianblue, urlcolor=oucrimsonred}
\definecolor{cornellred}{rgb}{0.7, 0.11, 0.11}
\definecolor{navyblue}{rgb}{0.0, 0.0, 0.5}
\definecolor{amethyst}{rgb}{0.6, 0.4, 0.8}
\definecolor{yellow}{rgb}{1.0, 1.0, 0.0}
\definecolor{firebrick}{rgb}{0.7, 0.13, 0.13}
\definecolor{tangerineyellow}{rgb}{1.0, 0.8, 0.0}
\definecolor{deepfuchsia}{rgb}{0.76, 0.33, 0.76}
\definecolor{amber}{rgb}{1.0, 0.75, 0.0}
\definecolor{VioletRed4}{rgb}{0.55, 0.13, .32}
\definecolor{indiagreen}{rgb}{0.07, 0.53, 0.03}
\definecolor{VioletRed4}{rgb}{0.55, 0.13, .32}

\usepackage{hyperref}
\usepackage{graphics, appendix, afterpage, makecell} 
\usepackage{bbold}
\usepackage{tikz}
\usepackage{adjustbox}

\usepackage{tcolorbox}

\definecolor{oucrimsonred}{rgb}{0.6, 0.0, 0.0}
\definecolor{persianblue}{rgb}{0.11, 0.22, 0.73}
\definecolor{forestgreen}{rgb}{0.13,0.35,0.13}
\definecolor{lightgray}{rgb}{0.83, 0.83, 0.83}
 \hypersetup{colorlinks, citecolor=oucrimsonred, linkcolor=persianblue, urlcolor=oucrimsonred}
\definecolor{cornellred}{rgb}{0.7, 0.11, 0.11}
\definecolor{navyblue}{rgb}{0.0, 0.0, 0.5}
\definecolor{amethyst}{rgb}{0.6, 0.4, 0.8}
\definecolor{yellow}{rgb}{1.0, 1.0, 0.0}
\definecolor{firebrick}{rgb}{0.7, 0.13, 0.13}
\definecolor{tangerineyellow}{rgb}{1.0, 0.8, 0.0}
\definecolor{deepfuchsia}{rgb}{0.76, 0.33, 0.76}
\definecolor{amber}{rgb}{1.0, 0.75, 0.0}
\definecolor{VioletRed4}{rgb}{0.55, 0.13, .32}
\definecolor{indiagreen}{rgb}{0.07, 0.53, 0.03}
\definecolor{VioletRed4}{rgb}{0.55, 0.13, .32}

\definecolor{oucrimsonred}{rgb}{0.6, 0.0, 0.0}
\newcommand\vertarrowbox[3][6ex]{%
  \begin{array}[t]{@{}c@{}} #2 \\
  \left\uparrow\vcenter{\hrule height #1}\right.\kern-\nulldelimiterspace\\
  \makebox[0pt]{\scriptsize#3}
  \end{array}%
}

\definecolor{mtcolor}{rgb}{.8,.3,.1}

\definecolor{violachiaro}{rgb}{1,0.6,1}

\definecolor{gbcolor}{rgb}{.43,.22,.12}
 
\definecolor{gbcolor2}{rgb}{.9,.2,.6}
\definecolor{gbcolor3}{rgb}{.3,.2,.6}

\definecolor{verdechiaro}{rgb}{0.6,1,0.6}
\definecolor{giallochiaro}{rgb}{1,1,0.6}
\definecolor{bluscuro}{rgb}{0.15, 0.2, 0.9}
\definecolor{verdes}{rgb}{0.1, 0.5, 0.1}%
\definecolor{tangerineyellow}{rgb}{1.0, 0.8, 0.0}
\definecolor{smokyblack}{rgb}{0.06, 0.05, 0.03}

\definecolor{americanrose}{rgb}{1.0, 0.01, 0.24}
\definecolor{cobalt}{rgb}{0.0, 0.28, 0.67}
\definecolor{brandeisblue}{rgb}{0.0, 0.44, 1.0}
\definecolor{mycolor}{rgb}{0.0, 0.0, 0.5}
\definecolor{oxfordblue}{rgb}{0.0, 0.13, 0.28}
\definecolor{azure}{rgb}{0.0, 0.5, 1.0}
\definecolor{turquoiseblue}{rgb}{0.0, 1.0, 0.94}
\newtcolorbox{mynewbox}[1]{colback=white!5!white,colframe=azure!75!black,fonttitle=\bfseries,title=#1}
\newtcolorbox{mybox}{colback=mycolor!5!white,colframe=azure!75!black}
\newtcolorbox{mynamedbox}[1]{colback=mycolor!5!white,colframe=azure!75!black,title=#1}
\definecolor{venetianred}{rgb}{0.78, 0.03, 0.08}
\newtcolorbox{mynamedbox1}[1]{colback=venetianred!5!white,colframe=venetianred!80!black,title=#1}
\newtcolorbox{mynamedbox2}[1]{colback=azure!5!white,colframe=azure!80!black,title=#1}

\definecolor{rossocorsa}{rgb}{0.83, 0.0, 0.0}

\tikzset{->-/.style={decoration={
  markings,
  mark=at position #1 with {\arrow{>}}},postaction={decorate}}}
\tikzset{-<-/.style={decoration={
  markings,
  mark=at position #1 with {\arrow{<}}},postaction={decorate}}} 

\def\be{\begin{equation}}
\def\ee{\end{equation}}
\def\ba{\begin{eqnarray}}
\def\ea{\end{eqnarray}}

\def\L*{{\cal L}_*}
\def\L{\mathcal{L}}
\def\({\left(}
\def\){\right)}

\def\<{\langle}
\def\>{\rangle}

 \def\neq {\not\equiv}

\def\cs2{c_{s}^{2}}

 \def\be   {\begin{equation}}   \def\ee   {\end{equation}}
 \def\ba   {\begin{array}}      \def\ea   {\end{array}}
 \def\bea  {\begin{eqnarray}}   \def\eea  {\end{eqnarray}}
 \def\bean {\begin{eqnarray*}}  \def\eean {\end{eqnarray*}}

\titleclass{\subsubsubsection}{straight}[\subsection]

\newcounter{subsubsubsection}[subsubsection]
\renewcommand\thesubsubsubsection{\thesubsubsection.\arabic{subsubsubsection}}

\titleformat{\subsubsubsection}
  {\normalfont\normalsize\bfseries}{\thesubsubsubsection}{1em}{}
\titlespacing*{\subsubsubsection}
{0pt}{3.25ex plus 1ex minus .2ex}{1.5ex plus .2ex}

\makeatletter
\renewcommand\paragraph{\@startsection{paragraph}{5}{\z@}%
  {3.25ex \@plus1ex \@minus.2ex}%
  {-1em}%
  {\normalfont\normalsize\bfseries}}
\renewcommand\subparagraph{\@startsection{subparagraph}{6}{\parindent}%
  {3.25ex \@plus1ex \@minus .2ex}%
  {-1em}%
  {\normalfont\normalsize\bfseries}}
\def\toclevel@subsubsubsection{4}
\def\toclevel@paragraph{5}
\def\toclevel@paragraph{6}
\def\l@subsubsubsection{\@dottedtocline{4}{7em}{4em}}
\def\l@paragraph{\@dottedtocline{5}{10em}{5em}}
\def\l@subparagraph{\@dottedtocline{6}{14em}{6em}}
\makeatother

\usepackage{titlesec} 
\usepackage[colorlinks,citecolor=blue,urlcolor=blue,hypertexnames=true]{hyperref}
\usepackage{subfigure}

\usepackage[usenames,dvipsnames,svgnames,table]{xcolor}

\usepackage{tikzsymbols}
\usepackage{natbib}
\usepackage{float}

\usepackage{tikz,xcolor,hyperref}

\usepackage{slashed}

\definecolor{lime}{HTML}{A6CE39}
\DeclareRobustCommand{\orcidicon}{
	\begin{tikzpicture}
	\draw[lime, fill=lime] (0,0) 
	circle [radius=0.2] 
	node[white] {{\fontfamily{qag}\selectfont \tiny ID}};
	\draw[white, fill=white] (-0.0625,0.095) 
	circle [radius=0.007];
	\end{tikzpicture}
	\hspace{-2mm}
}

\usepackage[usenames,dvipsnames,svgnames,table]{xcolor}  
\usepackage{tikzsymbols}
\usepackage{natbib}
\usepackage{float}

\usepackage{tikz,xcolor,hyperref}

\usepackage{slashed}

\definecolor{lime}{HTML}{A6CE39}
\DeclareRobustCommand{\orcidicon}{
	\begin{tikzpicture}
	\draw[lime, fill=lime] (0,0) 
	circle [radius=0.2] 
	node[white] {{\fontfamily{qag}\selectfont \tiny ID}};
	\draw[white, fill=white] (-0.0625,0.095) 
	circle [radius=0.007];
	\end{tikzpicture}
	\hspace{-2mm}
}

\foreach \x in {A, ..., Z}{\expandafter\xdef\csname orcid\x\endcsname{\noexpand\href{https://orcid.org/\csname orcidauthor\x\endcsname}
			{\noexpand\orcidicon}}
}
 
\usepackage{bbold}
\usepackage{tikz}
\usepackage{adjustbox}
\usepackage{tcolorbox}
\usepackage{enumitem}
\usepackage{amsfonts}

\setlist[itemize,1]{label=$\times$}
\setlist[itemize,2]{label=$\checkmark$}
\setlist[itemize,3]{label=$\diamond$}
\setlist[itemize,4]{label=$\bullet$}

\begin{document}
\title{\Large \textcolor{Sepia}{Symmetry-Protected $\alpha$-Attractor Hybrid Inflation in Supergravity and Constraints from ACT DR6 and DESI DR2}}
\author{\large Swapnil Kumar Singh\orcidF{}${}^{1}$}
\email{swapnilsingh.ph@gmail.com (Corresponding author)}

\affiliation{ ${}^{1}$B.M.S. College of Engineering, 
    Bangalore, Karnataka, 560019, India.}

\begin{abstract}
We present a symmetry-protected supergravity effective realization of hybrid $\alpha$-attractor inflation with a sequestered Stückelberg uplift. The model contains four chiral multiplets, an inflaton modulus $T$, a stabilizer field $S$, and a pair of oppositely charged waterfall multiplets $(\Psi,\bar\Psi)$, together with a hidden Stückelberg $U(1)_D$ sector that generates an approximately constant positive contribution to the scalar potential. Along the real inflationary trajectory, the F-term potential realizes the E-model plateau
\[
U(\phi)=V_0\left(1-e^{-\sqrt{2/(3\alpha)}\,\phi}\right)^2 ,
\]
while the sequestered uplift shifts the total energy density without directly modifying the tree-level inflaton slope. We derive the background evolution, the analytic $N$--$\phi$ relation, the waterfall stability condition, and the leading predictions for the scalar spectral index and tensor-to-scalar ratio. The model retains the characteristic red-tilted $\alpha$-attractor behavior, with $n_s\simeq1-2/N_*$ and tensor amplitudes controlled primarily by the curvature parameter $\alpha$, while the uplift and the hybrid end point generate only subleading corrections at fixed $N_*$. We also discuss the assumptions required for an effectively single-field trajectory, including the treatment of the light angular component of $T$, and examine the sensitivity of the sequestered uplift to Planck-suppressed cross-couplings. The resulting construction provides a controlled supergravity embedding of hybrid $\alpha$-attractor inflation compatible with current CMB constraints, including Planck, ACT DR6, and DESI DR2, while remaining testable by future $B$-mode searches such as LiteBIRD and CMB-S4.
\end{abstract}

\maketitle
\tableofcontents

\section{Introduction}

The inflationary paradigm provides a compelling description of the early Universe, offering a minimal and empirically successful mechanism for generating a spatially flat, homogeneous cosmos with a nearly Gaussian, adiabatic spectrum of scalar perturbations whose tilt is close to scale invariance~\cite{Guth:1981zm,Linde:1982ur,Albrecht:1982wi}. 
Precision measurements of the cosmic microwave background (CMB), especially by the \textit{Planck} satellite, have confirmed these predictions by tightly constraining the amplitude, spectral tilt, and tensor-to-scalar ratio of primordial perturbations~\cite{Planck:2018jri,Planck:2018vyg}. 

Forthcoming observations will further sharpen these tests. 
The Simons Observatory~\cite{SimonsObservatory:2019qwx} and the LiteBIRD satellite~\cite{LiteBIRD:2022cnt} aim to probe CMB $B$-modes with unprecedented sensitivity, while space-based gravitational-wave missions such as LISA, DECIGO, and BBO~\cite{Hild:2010id,Baker:2019nia,Smith:2019wny,Crowder:2005nr,Smith:2016jqs,Seto:2001qf,Kawamura:2020pcg,Bull:2018lat,LISACosmologyWorkingGroup:2022jok} target the stochastic background of inflationary gravitational waves. 
The recent confirmation of a stochastic signal by Pulsar Timing Array collaborations~\cite{NANOGrav:2023gor,Antoniadis:2023ott,Reardon:2023gzh,Xu:2023wog} has intensified interest in mechanisms that could yield such backgrounds, although canonical single-field slow-roll scenarios alone are unlikely to explain them~\cite{Vagnozzi:2023lwo,Oikonomou:2023qfz}. 
Meanwhile, recent analyses involving ACT DR6~\cite{ACT:2025fju,ACT:2025tim} and DESI DR2~\cite{DESI:2024uvr} have suggested a preference for a somewhat higher scalar spectral index than the central value inferred from \textit{Planck} alone. 
This has motivated renewed scrutiny of inflationary models in light of the broader observationally allowed region, including studies of modified plateau models, reheating effects, nonminimal couplings, higher-curvature corrections, Palatini scenarios, and related attractor constructions~\cite{Odintsov:2025bmp,Odintsov:2025jky,Odintsov:2025eiv,DOnofrio:2025bol,Ahmed:2025eip,Kim:2025dyi,Addazi:2025qra,Aoki:2025wld,Brahma:2025dio,Byrnes:2025kit,Choudhury:2025vso,Dioguardi:2025mpp,Dioguardi:2025vci,Drees:2025ngb,Ferreira:2025lrd,Frolovsky:2025iao,Gao:2025onc,Gao:2025viy,Gialamas:2025kef,Gialamas:2025ofz,Hai:2025wvs,Kallosh:2025rni,Kouniatalis:2025orn,Liu:2025qca,Mohammadi:2025gbu,Odintsov:2025wai,Peng:2025bws,Q:2025ycf,Salvio:2025izr,Wolf:2025ecy,Yin:2025rrs,Yi:2025dms,Yogesh:2025wak,Yuennan:2025kde,Zahoor:2025nuq,Zhu:2025twm,aoki2025higgsmodularinflation,chakraborty2025nonminimalinfraredgravitationalreheating,mondal2025constrainingreheatingtemperatureinflatonsm,pallis2025kineticallymodifiedpalatiniinflation,wolf2025inflationaryattractorsradiativecorrections,Oikonomou:2025xms,Oikonomou:2025htz}. 
The construction studied in this work should be viewed in this context. 
It remains a red-tilted $\alpha$-attractor model, typically predicting $n_s\simeq0.967$--$0.968$ for $N_*=60$, and is therefore compatible with the newer ACT DR6 and DESI DR2 constraints without being specifically engineered to reproduce the central value of the ACT-preferred higher-$n_s$ region. 
This distinction is important: the model preserves the standard attractor mechanism rather than introducing a deformation designed to generate a blue-shifted scalar tilt.

Hybrid inflation~\cite{Linde:1993cn} occupies a special place in this landscape. 
Its dynamics proceed along a quasi-flat valley in a multi-field potential and terminate when an orthogonal ``waterfall'' field becomes tachyonic, ending inflation abruptly. 
This built-in exit mechanism avoids ambiguities in slow-roll termination and enables controlled symmetry breaking in the post-inflationary phase. 
Embedding hybrid inflation in supergravity (SUGRA) introduces the familiar $\eta$-problem: order-one corrections from the K\"ahler curvature can spoil the flatness of the inflaton potential~\cite{Copeland:1994vg,Dvali:1994ms,Binetruy:1996xj,Halyo:1996pp}. 
The framework of $\alpha$-attractors provides a geometric way to control this problem~\cite{Kallosh:2013yoa,Kallosh:2013hoa,Ferrara:2013rsa,Galante:2014ifa,Linde:2015uga}. 
In these models the inflaton spans a hyperbolic manifold, usually represented by the Poincar\'e disk or half-plane, whose negative curvature stretches the canonical field range near the boundary and flattens a wide class of potentials after canonical normalization. 
The resulting E- and T-models predict
\[
n_s\simeq 1-\frac{2}{N_*}, 
\qquad 
r\simeq \mathcal{O}\!\left(\frac{\alpha}{N_*^2}\right),
\]
exhibiting the characteristic universality of the attractor regime.

Recent works~\cite{Pallis:2025gii} have analyzed the interplay between geometric stabilization, attractor dynamics, and F- and D-term contributions in related supergravity settings. 
These developments highlight the importance of models in which uplift sectors, stabilizer fields, and waterfall dynamics are sufficiently controlled to preserve the attractor structure while allowing a viable post-inflationary vacuum.

In this work we construct a four-dimensional $\mathcal{N}=1$ supergravity effective realization that merges hybrid inflation with $\alpha$-attractor geometry. 
The model contains four chiral multiplets,
\[
(T,S,\Psi,\bar\Psi),
\]
where $T$ is the inflaton modulus, $S$ is a stabilizer field, and $(\Psi,\bar\Psi)$ are oppositely charged waterfall multiplets under a gauged $U(1)_X$. 
In addition, a hidden St\"uckelberg $U(1)_D$ sector generates an approximately constant positive uplift along the inflationary valley. 
This construction should be understood as a controlled supergravity embedding rather than a UV-complete theory by itself. 
The uplift sector is assumed to be sequestered from the inflaton sector at the level of the two-derivative effective theory, and we explicitly discuss below the sensitivity of this assumption to Planck-suppressed cross-couplings.

Along the real inflationary trajectory, the stabilizer F-term potential realizes the E-model plateau
\[
U(\phi)=V_0\left(1-e^{-\beta\phi}\right)^2,
\qquad 
\beta=\sqrt{\frac{2}{3\alpha}},
\]
while the St\"uckelberg sector contributes an approximately constant uplift to the total scalar potential. 
The inflaton corresponds to the canonically normalized real part of $T$ on the half-plane, and the $T$-dependent waterfall mass controls the hybrid transition. 
We derive the background dynamics, the analytic $N$--$\phi$ relation in the presence of the uplift, the waterfall stability condition, and the leading predictions for $(n_s,r)$. 
The model preserves the characteristic red-tilted $\alpha$-attractor prediction for $n_s$, while the tensor amplitude is controlled primarily by the curvature parameter $\alpha$ and receives only subleading corrections from the uplift and the precise hybrid end point at fixed $N_*$. 
Next-to-leading-order predictions are computed using the Hubble-flow hierarchy~\cite{Stewart:1993zq,Leach:2002ar,Gong:2001he,Schwarz:2001vv}, and quantum stability is examined through Coleman--Weinberg corrections~\cite{Coleman:1973jx,Jackiw:1974cv,Martin:2001vx}.

A further subtlety is that, in the minimal K\"ahler realization, the angular component of the modulus $T$ can remain lighter than the Hubble scale during inflation. 
Therefore, the single-field predictions quoted in this work should be interpreted as applying to the stabilized or trajectory-aligned branch of the model. 
Equivalently, one may assume that initial conditions select the real trajectory with negligible turn rate, or that small higher-order K\"ahler corrections stabilize the angular mode without significantly modifying the real-axis plateau. 
We include an estimate of the possible isocurvature contribution and specify the assumptions under which it remains negligible.

We adopt reduced Planck units $M_{\rm Pl}=1$ unless otherwise stated.\footnote{Canonical scalar fields have mass dimension~1; the holomorphic superpotential $W$ has dimension~3; parameters such as $V_0=\mu^2$ carry dimension~4; the Yukawa-like parameter $g$ in $W$ has dimension~1; and the gauge couplings $g_X$, $g_D$ and quartic coefficients are dimensionless.}

The remainder of this paper is organized as follows. 
In Section~\ref{sec:model}, we present the supergravity construction, including the K\"ahler potential, superpotential, gauge sectors, waterfall dynamics, and the sequestered St\"uckelberg uplift mechanism. 
We also discuss the sensitivity of the sequestered structure to Planck-suppressed cross-couplings. 
Section~\ref{sec:background} derives the background dynamics and the analytic $N$--$\phi$ mapping in the presence of a constant uplift. 
In Section~\ref{sec:perturbations}, we compute the inflationary perturbations using the Hubble-flow hierarchy and discuss the conditions under which the effectively single-field treatment is valid. 
Section~\ref{sec:postinflation} discusses reheating, stabilization, topological defects, and vacuum structure. 
In Section~\ref{sec:obs}, we compare the theoretical predictions with current CMB constraints, including Planck, ACT DR6, and DESI DR2, and analyze the dependence on model parameters. 
Finally, Section~\ref{sec:conclusion} summarizes our results and outlines possible extensions of the framework.
\section{Supergravity construction}
\label{sec:model}

\subsection*{Field content, K\"ahler geometry, and superpotential}

We begin by specifying the chiral and gauge content of the inflationary sector. The model contains four chiral multiplets,
\[
(T,S,\Psi,\bar\Psi),
\]
where $T$ is the inflaton modulus, $S$ is a stabilizer field, and $\Psi$ and $\bar\Psi$ are waterfall multiplets with opposite charges $+1$ and $-1$ under an Abelian gauge group $U(1)_X$. The modulus $T$ parametrizes the Poincar\'e half-plane, whose curvature is controlled by the positive parameter $\alpha$~\cite{Kallosh:2013yoa,Roest:2013fha,Kallosh:2015zsa}. This hyperbolic geometry is the origin of the attractor behavior after canonical normalization.

The K\"ahler potential of the inflationary sector is chosen as
\begin{align}
K_{\rm infl}
=
-3\alpha\,\ln(T+\bar T)
+
(T+\bar T)^{-3\alpha}
\left(
|S|^2+|\Psi|^2+|\bar\Psi|^2
\right).
\label{eq:K}
\end{align}
This form makes the matter-field kinetic metrics scale uniformly with the real part of $T$. The common prefactor $(T+\bar T)^{-3\alpha}$ is useful because it leads to a cancellation in the stabilizer F-term energy along the inflationary valley. At the same time, the matter fields are not completely independent of the inflaton geometry: their kinetic normalization is $T$-dependent and must be included when physical masses are discussed.\footnote{The phrase ``physical mass'' will always refer to the mass after canonical normalization of the corresponding scalar fluctuation. Coefficients appearing directly in the scalar potential before rescaling by the K\"ahler metric are coordinate-basis masses.}

Along the real inflationary valley
\[
S=\Psi=\bar\Psi=0,
\qquad
T=\bar T>0,
\]
the nonvanishing components of the K\"ahler metric are
\begin{align}
K_{T\bar T}
=
\frac{3\alpha}{(T+\bar T)^2},
\qquad
K_{S\bar S}
=
K_{\Psi\bar\Psi}
=
K_{\bar\Psi\overline{\bar\Psi}}
=
(T+\bar T)^{-3\alpha}.
\label{eq:Kmetric_components}
\end{align}
The corresponding inverse components are
\begin{align}
K^{S\bar S}
=
K^{\Psi\bar\Psi}
=
K^{\bar\Psi\overline{\bar\Psi}}
=
(T+\bar T)^{3\alpha},
\label{eq:Kmetric_inverse_components}
\end{align}
and the exponential factor reduces to
\[
e^K=(T+\bar T)^{-3\alpha}.
\]
Therefore,
\begin{align}
e^K K^{S\bar S}=1,
\qquad
e^K K^{\Psi\bar\Psi}
=
e^K K^{\bar\Psi\overline{\bar\Psi}}
=
1 .
\label{eq:EK_cancellation}
\end{align}
The first of these relations is responsible for the exact cancellation of the conformal prefactor in the stabilizer F-term energy density. The second relation is useful for identifying the coordinate-basis waterfall mass terms, while the physical masses still require division by the matter-field kinetic metric.

The inflationary-sector holomorphic superpotential is taken to be
\begin{align}
W_{\rm infl}
=
S\,\mu\,(1-T^{-1})
+
g\,(1-T^{-1})\,\Psi\,\bar\Psi .
\label{eq:W}
\end{align}
This superpotential is invariant under $U(1)_X$ because the product $\Psi\bar\Psi$ is neutral. 
The parameter $\mu$ sets the inflationary energy scale, while $g$ controls the $T$-dependent supersymmetric mass of the waterfall pair. 
The factor $(1-T^{-1})$ fixes the supersymmetric point of the inflaton sector at $T=1$, where the F-term contribution from the stabilizer vanishes.

The St\"uckelberg multiplet $\Sigma$ is not included in the perturbative holomorphic superpotential of the sequestered inflationary effective theory. 
Thus the superpotential relevant during inflation is
\begin{align}
W_{\rm tot}
=
W_{\rm infl}(T,S,\Psi,\bar\Psi),
\qquad
W_{\rm Stk}(\Sigma)=0,
\qquad
\partial_\Sigma W_{\rm tot}=0 .
\label{eq:W_no_Stk}
\end{align}
This is a symmetry statement rather than an additional fine-tuning assumption: the St\"uckelberg sector gauges the axionic shift symmetry
\[
\delta\Sigma=iq_D\Lambda_D,
\]
and a perturbative holomorphic dependence on $\Sigma$ would violate this gauged shift symmetry unless accompanied by additional charged spurions or explicit shift-breaking effects. 
Such effects are not part of the sequestered branch considered here and are assumed to be absent, or exponentially suppressed, over the inflationary regime. 
Consequently, the hidden St\"uckelberg sector does not generate an independent F-term contribution during inflation. 
Its distinctive role in the construction is to provide the approximately constant D-term uplift described below.

\subsection*{F-term potential and the inflationary valley}

In $\mathcal{N}=1$ supergravity, the F-term scalar potential is given by~\cite{Wess:1992cp,Freedman:2012zz}
\begin{align}
V_F
=
e^K
\left(
K^{I\bar J}D_IW_{\rm tot}D_{\bar J}\overline W_{\rm tot}
-
3|W_{\rm tot}|^2
\right),
\qquad
D_IW_{\rm tot}=\partial_IW_{\rm tot}+K_IW_{\rm tot},
\label{eq:VFgeneral}
\end{align}
where $I$ and $\bar J$ run over the chiral multiplets present in the effective theory. 
For the inflationary sector the only nonzero F-term on the valley is the stabilizer F-term. 
Including the sequestered St\"uckelberg multiplet, Eq.~\eqref{eq:W_no_Stk} gives
\begin{align}
W_{\rm tot}\big|_{\rm val}=0,
\qquad
D_T W_{\rm tot}\big|_{\rm val}=0,
\qquad
D_S W_{\rm tot}\big|_{\rm val}
=
\mu(1-T^{-1}),
\qquad
D_\Sigma W_{\rm tot}\big|_{\rm val}=0 .
\label{eq:valley_data}
\end{align}
The last equality follows from
\begin{align}
D_\Sigma W_{\rm tot}\big|_{\rm val}
=
\partial_\Sigma W_{\rm tot}\big|_{\rm val}
+
K_\Sigma W_{\rm tot}\big|_{\rm val}
=
0 ,
\end{align}
because $\partial_\Sigma W_{\rm tot}=0$ and $W_{\rm tot}\big|_{\rm val}=0$. 
Therefore the St\"uckelberg multiplet does not contribute to $V_F$ along the inflationary trajectory. 
The vanishing of $W_{\rm tot}$ also removes the negative supergravity term $-3e^K|W_{\rm tot}|^2$. 
Thus the F-term valley energy is generated entirely by the stabilizer multiplet:
\begin{align}
V_F\big|_{\rm val}
=
e^K K^{S\bar S}|D_SW_{\rm tot}|^2
=
\mu^2|1-T^{-1}|^2 .
\label{eq:VFvalley}
\end{align}

This is an exact tree-level result within the two-derivative supergravity effective theory specified by Eqs.~\eqref{eq:K} and \eqref{eq:W}. It is not a statement of ultraviolet completeness; possible higher-dimensional operators are assumed to be under perturbative control in the effective description.\footnote{Throughout this work, ``supergravity embedding'' means a controlled four-dimensional $\mathcal{N}=1$ effective realization. It does not imply that the finite set of fields in Eq.~\eqref{eq:W} constitutes a UV-complete theory.}

The quadratic dependence of $V_F$ on the waterfall fields can also be obtained explicitly. Let
\[
x=T+\bar T,
\qquad
A(T)=1-T^{-1},
\qquad
q(x)=x^{-3\alpha}.
\]
Keeping terms up to quadratic order in $\Psi$ and $\bar\Psi$ at $S=0$, one finds
\begin{align}
e^K K^{S\bar S}|D_SW_{\rm tot}|^2
=
\mu^2|A|^2
\left[
1+q(x)\left(|\Psi|^2+|\bar\Psi|^2\right)
+\mathcal{O}(|\Psi|^4,|\bar\Psi|^4)
\right].
\label{eq:Fterm_waterfall_expansion}
\end{align}
The first term is the inflaton potential on the valley, while the second term gives a positive Hubble-scale contribution to the canonically normalized waterfall masses. Since the kinetic term of $\Psi$ is $q(x)\partial\Psi\partial\bar\Psi$, the canonically normalized fields are locally $\Psi_c=q^{1/2}\Psi$ and $\bar\Psi_c=q^{1/2}\bar\Psi$. Hence the stabilizer F-term contributes
\begin{align}
\Delta m_H^2(\phi)=U(\phi)
\label{eq:Hubble_mass_minimal}
\end{align}
to both waterfall eigenvalues in the minimal K\"ahler realization. More general higher-order K\"ahler operators can change the coefficient of this contribution, and below we write it as $c_HU(\phi)$ with $c_H=\mathcal{O}(1)$ in the absence of special cancellations.

\subsection*{Sequestered St\"uckelberg uplift and its naturalness}

To generate an approximately constant positive contribution during inflation, we add a hidden $U(1)_D$ gauge sector coupled to a St\"uckelberg chiral multiplet $\Sigma$. The sector gauges the axionic shift symmetry
\begin{align}
\delta\Sigma=i q_D\Lambda_D,
\qquad
K_{\rm Stk}=k(\Sigma+\bar\Sigma+q_DV_D),
\qquad
f_D={\rm constant}.
\label{eq:Stk_data}
\end{align}
The corresponding Killing potential is
\begin{align}
\mathcal{P}_D
=
-q_D\,\partial_{\Sigma+\bar\Sigma}k
\big|_{\langle\Sigma\rangle}
\equiv
\xi_{\rm eff},
\label{eq:PD_def}
\end{align}
where $\xi_{\rm eff}$ is constant after the heavy St\"uckelberg multiplet and the $U(1)_D$ vector are integrated out. The D-term scalar potential then gives
\begin{align}
V_D
=
\frac{g_D^2}{2}\xi_{\rm eff}^2
\equiv
V_{\rm up}.
\label{eq:Vup}
\end{align}

The sequestered effective structure used in this construction is
\begin{align}
K
&=
K_{\rm infl}(T,S,\Psi,\bar\Psi)
+
K_{\rm Stk}(\Sigma+\bar\Sigma+q_DV_D),
\nonumber\\
f_{AB}
&=
{\rm diag}\left(f_X,f_D\right),
\nonumber\\
W_{\rm tot}
&=
W_{\rm infl}(T,S,\Psi,\bar\Psi),
\qquad
\partial_\Sigma W_{\rm tot}=0 .
\label{eq:factorized_K}
\end{align}
Thus the hidden St\"uckelberg sector enters through the K\"ahler potential, the gauged shift symmetry, and the $U(1)_D$ gauge dynamics, but not through a separate $\Sigma$-dependent superpotential. 
The uplift in Eq.~\eqref{eq:Vup} is therefore a genuinely D-term St\"uckelberg uplift rather than an F-term uplift in disguise. 
This is the sequestered branch of the effective theory. 
In this branch, integrating out the heavy St\"uckelberg multiplet and the $U(1)_D$ vector adds an approximately constant positive contribution to the scalar potential and does not generate an additional tree-level force along the inflaton direction.

The effective valley potential then has the form
\begin{align}
V_{\rm val}(\phi)
=
U(\phi)+V_{\rm up},
\label{eq:Vval_up}
\end{align}
up to the constant false-vacuum contribution from the waterfall D-term discussed below.

The factorization in Eq.~\eqref{eq:factorized_K} is an assumption of the effective supergravity construction. In a generic supergravity theory, Planck-suppressed cross-couplings between the inflaton and St\"uckelberg sectors may be present. 

A Planck-suppressed correction to the sequestered structure is
\begin{align}
\Delta K
=
\epsilon_\Sigma\,
\mathcal{F}_\Sigma(T+\bar T)
\left(
\Sigma+\bar\Sigma+q_DV_D
\right)^2 ,
\label{eq:DeltaK_cross}
\end{align}
where $\epsilon_\Sigma$ is a dimensionless coefficient and 
$\mathcal{F}_\Sigma(x)$, with $x=T+\bar T$, is a real dimensionless function parametrizing the leading K\"ahler contact interaction between the inflaton modulus and the St\"uckelberg sector. 
The function $\mathcal{F}_\Sigma$ is not part of the sequestered tree-level model; it is introduced only to quantify possible departures from exact sequestering. 
Along the real inflationary trajectory $T=\bar T=e^{\beta\phi}$, we define
\begin{align}
\widehat{\mathcal{F}}_\Sigma(\phi)
\equiv
\mathcal{F}_\Sigma(2e^{\beta\phi}) .
\label{eq:Fhat_def}
\end{align}
After the heavy St\"uckelberg multiplet and the $U(1)_D$ vector are integrated out, such a term generically induces a weak inflaton dependence in the uplift,
\begin{align}
V_{\rm up}
\longrightarrow
V_{\rm up}(\phi)
=
V_{\rm up}
\left[
1+
\epsilon_\Sigma
\widehat{\mathcal{F}}_\Sigma(\phi)
+
\mathcal{O}(\epsilon_\Sigma^2)
\right].
\label{eq:Vup_phi}
\end{align}
The induced correction to the slow-roll curvature is parametrically
\begin{align}
\Delta\eta_{\rm up}
\sim
\frac{V_{\rm up}}{V}\,
\epsilon_\Sigma\,
\frac{d^2\widehat{\mathcal{F}}_\Sigma}{d\phi^2}.
\label{eq:Delta_eta_up}
\end{align}
If $\widehat{\mathcal{F}}_\Sigma$ and its first few $\phi$-derivatives vary by order unity over the observable inflationary range, preservation of the attractor plateau requires
\begin{align}
|\epsilon_\Sigma|
\frac{V_{\rm up}}{V}
\lesssim
10^{-2}.
\label{eq:epsSigma_bound}
\end{align}
A stronger restriction is required if $\widehat{\mathcal{F}}_\Sigma$ grows rapidly near the boundary of the half-plane.

Thus, the constant uplift is technically stable in the sequestered branch of the effective theory, where such operators are absent or sufficiently suppressed. This can occur due to the gauged shift symmetry, locality in an extra-dimensional realization, a no-scale structure, or charge assignments that forbid direct $T$-$\Sigma$ mixing. If unsuppressed cross-couplings are allowed, the uplift is no longer constant and the attractor flatness can be spoiled.

\subsubsection*{One-loop K\"ahler renormalization of the sequestered uplift}

The previous estimate treated the coefficient $\epsilon_\Sigma$ as a generic effective-theory parameter. 
We now estimate the corresponding contribution induced by the logarithmically divergent one-loop K\"ahler renormalization of supergravity. 
This is the relevant quantum check of the constant-uplift assumption: if the one-loop K\"ahler correction generates an unsuppressed $T$--$\Sigma$ contact term, the uplift would acquire an inflaton dependence and could spoil the attractor plateau. 
We show below that this does not happen in the exactly sequestered branch, and that even a mildly non-sequestered threshold contribution is loop-suppressed.

The logarithmically divergent part of the one-loop effective action contains terms that can be absorbed into a renormalized K\"ahler potential,
\begin{align}
K_R
=
K+\Delta K_{1{\rm -loop}} .
\end{align}
The part relevant for the K\"ahler renormalization may be written, following Eq.~(4.10) of Ref.~\cite{Gaillard:1996ms}, as
\begin{align}
\Delta K_{1{\rm -loop}}
=
\frac{\mathcal L_\Lambda}{32\pi^2}
\left[
e^{-K}
\left(
A_{ij}\bar A^{ij}
-
2A_i\bar A^i
-
4A\bar A
\right)
-
4K^a{}_{a}
-
\left(12+4x^2\rho_i\rho^i\right)\mathcal D
\right],
\label{eq:DeltaK_1loop_general}
\end{align}
where
\begin{align}
\mathcal L_\Lambda
\equiv
\ln\left(\frac{\Lambda^2}{\mu_R^2}\right).
\end{align}
Here $i,j$ run over the chiral multiplets, $A,A_i,A_{ij}$ denote the K\"ahler-covariant superpotential quantities in the conventions of Ref.~\cite{Gaillard:1996ms}, $K^a{}_{a}$ denotes the trace over gauge Killing directions, $\mathcal D$ denotes the D-term auxiliary contribution in the same conventions, $x={\rm Re}\,f$, and
\begin{align}
\rho_i
\equiv
\frac{\partial_i x}{2x}.
\label{eq:rho_def}
\end{align}
In the present construction the gauge kinetic functions of both Abelian sectors are taken to be field-independent,
\begin{align}
f_X={\rm constant},
\qquad
f_D={\rm constant},
\end{align}
and hence
\begin{align}
\rho_i=0 .
\label{eq:rho_zero}
\end{align}
Therefore the terms in Eq.~\eqref{eq:DeltaK_1loop_general} involving derivatives of the gauge kinetic function do not induce an inflaton--St\"uckelberg mixing.

The sequestered branch of the effective theory is defined by
\begin{align}
K
=
K_{\rm infl}(T,S,\Psi,\bar\Psi)
+
K_{\rm Stk}(Y),
\qquad
Y\equiv\Sigma+\bar\Sigma+q_DV_D,
\label{eq:sequestered_K_loop}
\end{align}
together with
\begin{align}
W_{\rm tot}
=
W_{\rm infl}(T,S,\Psi,\bar\Psi),
\qquad
\partial_\Sigma W_{\rm tot}=0 .
\label{eq:sequestered_W_loop}
\end{align}
Along the inflationary valley,
\begin{align}
S=\Psi=\bar\Psi=0,
\qquad
W_{\rm tot}\big|_{\rm val}=0,
\qquad
D_\Sigma W_{\rm tot}\big|_{\rm val}=0,
\qquad
D_S W_{\rm tot}\big|_{\rm val}
=
\mu(1-T^{-1}) .
\label{eq:loop_valley_conditions}
\end{align}
Thus the St\"uckelberg multiplet has no F-term on the inflationary trajectory. 
The inflaton plateau is generated by the stabilizer F-term, while the hidden $U(1)_D$ sector contributes through its Killing potential and D-term uplift.

To isolate possible one-loop $T$--$\Sigma$ mixing, expand the correction around the heavy St\"uckelberg minimum $Y=Y_0$:
\begin{align}
\Delta K_{1{\rm -loop}}
=
\Delta K_{1{\rm -loop}}^{\rm infl}(T+\bar T)
+
\Delta k_{1{\rm -loop}}^{\rm Stk}(Y)
+
\frac{1}{2}
C_{TYY}^{(1)}(T+\bar T)
\left(Y-Y_0\right)^2
+\cdots .
\label{eq:DeltaK_loop_expansion}
\end{align}
The genuinely mixed part is measured by
\begin{align}
\left.
\frac{\partial}{\partial(T+\bar T)}
\frac{\partial^2\Delta K_{1{\rm -loop}}}
{\partial Y^2}
\right|_{\rm val}.
\label{eq:mixed_derivative_detector}
\end{align}
In the strictly factorized effective theory, with no heavy messenger state coupling simultaneously to the inflaton and St\"uckelberg sectors, the one-loop correction separates into an inflationary contribution and a hidden-sector contribution:
\begin{align}
\Delta K_{1{\rm -loop}}\big|_{\rm val}
=
\Delta K_{1{\rm -loop}}^{\rm infl}(T+\bar T)
+
\Delta k_{1{\rm -loop}}^{\rm Stk}(Y)
+
{\rm constant}.
\label{eq:DeltaK_1loop_factorized}
\end{align}
Consequently,
\begin{align}
\left.
\frac{\partial}{\partial(T+\bar T)}
\frac{\partial^2\Delta K_{1{\rm -loop}}}
{\partial Y^2}
\right|_{\rm val}
=
0 .
\label{eq:no_loop_mixing_exact}
\end{align}
Thus, in the exactly sequestered branch, the St\"uckelberg sector renormalizes the hidden-sector K\"ahler function and the constant uplift, but it does not generate an additional one-loop inflaton force.

If sequestering is mildly violated by heavy threshold states, a field-dependent regulator, or messenger fields communicating between the two sectors, the induced mixed operator has the same structure as the Planck-suppressed diagnostic operator introduced above:
\begin{align}
\Delta K_{1{\rm -loop}}^{\rm mix}
=
\epsilon_\Sigma^{(1)}
\mathcal F_\Sigma^{(1)}(T+\bar T)
\left(Y-Y_0\right)^2 .
\label{eq:DeltaK_loop_mix}
\end{align}
The natural size of the coefficient is fixed by the one-loop factor,
\begin{align}
\epsilon_\Sigma^{(1)}
\sim
\frac{c_{\rm mix}}{32\pi^2}
\ln\left(\frac{\Lambda^2}{M_{\rm Stk}^2}\right),
\label{eq:eps_loop_estimate}
\end{align}
where $M_{\rm Stk}$ denotes the heavy St\"uckelberg/vector mass scale. 
The dimensionless coefficient $c_{\rm mix}$ measures the strength of non-sequestered threshold communication. 
It vanishes in the exactly factorized theory,
\begin{align}
c_{\rm mix}=0,
\label{eq:cmix_zero}
\end{align}
whereas $c_{\rm mix}=\mathcal O(1)$ would correspond to an unsuppressed heavy threshold coupling to both sectors.

After integrating out the heavy St\"uckelberg multiplet, Eq.~\eqref{eq:DeltaK_loop_mix} induces
\begin{align}
V_{\rm up}
\longrightarrow
V_{\rm up}(\phi)
=
V_{\rm up}
\left[
1+
\epsilon_\Sigma^{(1)}
\widehat{\mathcal F}_\Sigma^{(1)}(\phi)
+
\mathcal O\!\left((\epsilon_\Sigma^{(1)})^2\right)
\right],
\label{eq:Vup_loop_phi}
\end{align}
where
\begin{align}
\widehat{\mathcal F}_\Sigma^{(1)}(\phi)
\equiv
\mathcal F_\Sigma^{(1)}(2e^{\beta\phi}) .
\end{align}
The induced correction to the slow-roll curvature is
\begin{align}
\Delta\eta_{\rm up}^{(1)}
\simeq
\frac{V_{\rm up}}{V}
\epsilon_\Sigma^{(1)}
\frac{d^2\widehat{\mathcal F}_\Sigma^{(1)}}{d\phi^2}.
\label{eq:Delta_eta_loop_general}
\end{align}
Using Eq.~\eqref{eq:eps_loop_estimate}, we obtain
\begin{align}
|\Delta\eta_{\rm up}^{(1)}|
\sim
\frac{V_{\rm up}}{V}
\frac{|c_{\rm mix}|}{32\pi^2}
\ln\left(\frac{\Lambda^2}{M_{\rm Stk}^2}\right)
\left|
\frac{d^2\widehat{\mathcal F}_\Sigma^{(1)}}{d\phi^2}
\right|.
\label{eq:Delta_eta_loop_estimate}
\end{align}
This expression is the useful quantitative result: the one-loop correction is either exactly absent in the sequestered branch, or else it is suppressed by the universal factor $1/(32\pi^2)$.

It is useful to separate the part of Eq.~\eqref{eq:Delta_eta_loop_estimate} that is fixed by the effective theory from the part that depends on unknown UV threshold data. 
Define
\begin{align}
R
\equiv
\frac{V_{\rm up}}{V},
\qquad
L
\equiv
\ln\left(\frac{\Lambda^2}{M_{\rm Stk}^2}\right),
\qquad
Q
\equiv
|c_{\rm mix}|
\left|
\frac{d^2\widehat{\mathcal F}_\Sigma^{(1)}}{d\phi^2}
\right|.
\end{align}
Then
\begin{align}
|\Delta\eta_{\rm up}^{(1)}|
=
Q\,\Xi,
\qquad
\Xi
\equiv
\frac{RL}{32\pi^2}.
\label{eq:Xi_definition}
\end{align}
The quantity $\Xi$ is the exact normalized one-loop correction fixed by the threshold logarithm and uplift fraction. 
The model-dependent factor $Q$ is zero in the exactly sequestered branch and is at most order unity only if non-sequestered heavy thresholds couple directly to both sectors.

The numerical boundary for a percent-level curvature correction is obtained exactly from
\begin{align}
Q\,\Xi
=
10^{-2}.
\end{align}
For $Q=1$ this gives
\begin{align}
RL
=
32\pi^2\times10^{-2}
=
3.158273\ldots .
\label{eq:percent_boundary}
\end{align}
Similarly, the $10^{-3}$ contour is
\begin{align}
RL
=
32\pi^2\times10^{-3}
=
0.315827\ldots .
\label{eq:permille_boundary}
\end{align}
These numerical values are the exact contour values following from Eq.~\eqref{eq:Xi_definition}.

\begin{table}[t]
\centering
\renewcommand{\arraystretch}{1.35}
\begin{tabular}{c c c}
\toprule
Condition & Exact relation & Numerical value for $Q=1$ \\
\midrule
One-loop normalization 
& $32\pi^2$ 
& $315.8273408$ \\
\midrule
$|\Delta\eta_{\rm up}^{(1)}|=10^{-3}$ 
& $RL=32\pi^2\times10^{-3}$ 
& $0.3158273$ \\
$|\Delta\eta_{\rm up}^{(1)}|=10^{-2}$ 
& $RL=32\pi^2\times10^{-2}$ 
& $3.1582734$ \\
\midrule
$R=1.00$ boundary 
& $L_c=3.1582734/R$ 
& $L_c=3.1583$ \\
$R=0.75$ boundary 
& $L_c=3.1582734/R$ 
& $L_c=4.2110$ \\
$R=0.50$ boundary 
& $L_c=3.1582734/R$ 
& $L_c=6.3165$ \\
$R=0.25$ boundary 
& $L_c=3.1582734/R$ 
& $L_c=12.6331$ \\
$R=0.10$ boundary 
& $L_c=3.1582734/R$ 
& $L_c=31.5827$ \\
\midrule
$L=1$ boundary 
& $R_c=3.1582734/L$ 
& $R_c=3.1583$ \\
$L=3$ boundary 
& $R_c=3.1582734/L$ 
& $R_c=1.0528$ \\
$L=5$ boundary 
& $R_c=3.1582734/L$ 
& $R_c=0.6317$ \\
$L=8$ boundary 
& $R_c=3.1582734/L$ 
& $R_c=0.3948$ \\
$L=10$ boundary 
& $R_c=3.1582734/L$ 
& $R_c=0.3158$ \\
\bottomrule
\end{tabular}
\caption{
Exact numerical thresholds for the normalized one-loop uplift-curvature correction.
Here $Q\equiv |c_{\rm mix}|\left|d^2\widehat{\mathcal F}^{(1)}_\Sigma/d\phi^2\right|$ and
$|\Delta\eta_{\rm up}^{(1)}|=Q\,RL/(32\pi^2)$.
The table gives the exact $Q=1$ threshold values. 
For general $Q$, the critical products scale as $RL\rightarrow RL/Q$. 
In the exactly sequestered branch, $Q=0$, so the mixed one-loop correction vanishes identically.}
\label{tab:loop_eta_exact_thresholds}
\end{table}

\begin{figure}[t]
\centering
\includegraphics[width=0.86\textwidth]{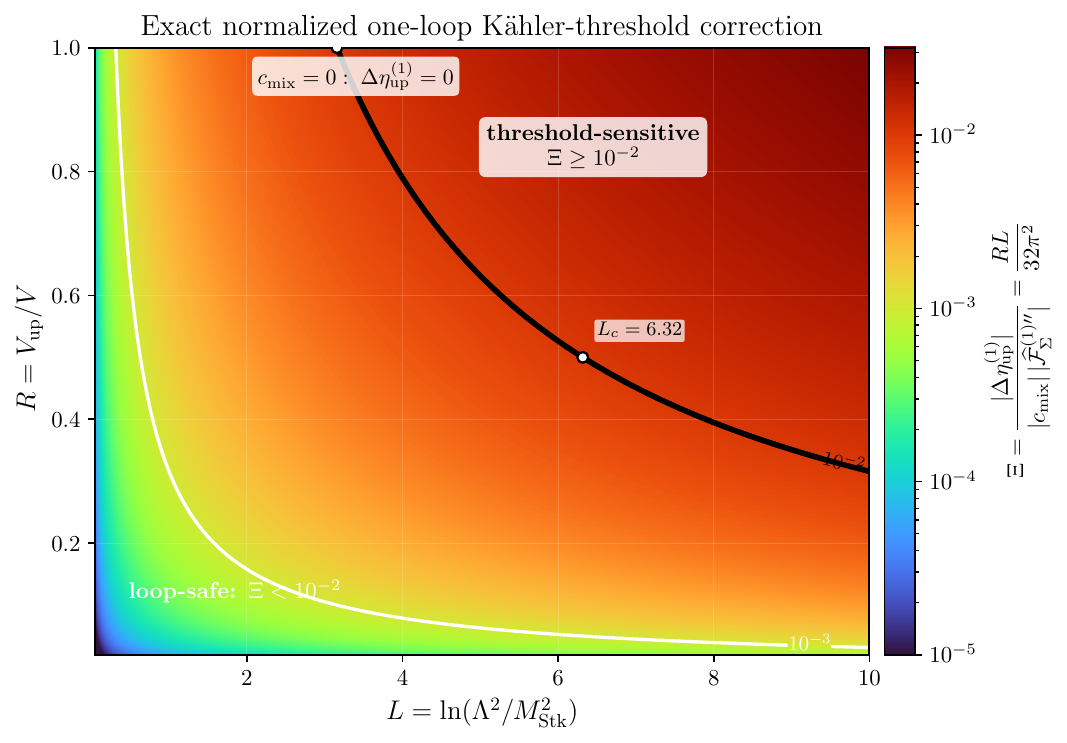}
\caption{
Exact normalized one-loop K\"ahler-threshold correction to the sequestered St\"uckelberg uplift.
The color map shows 
$\Xi=|\Delta\eta_{\rm up}^{(1)}|/
\left(|c_{\rm mix}|\,|\widehat{\mathcal F}_{\Sigma}^{(1)\prime\prime}|\right)
=RL/(32\pi^2)$, where 
$R=V_{\rm up}/V$ and 
$L=\ln(\Lambda^2/M_{\rm Stk}^2)$.
The white contour is $\Xi=10^{-3}$ and the black contour is $\Xi=10^{-2}$.
The black contour is the percent-level slow-roll boundary for $Q=1$.
Below it, the loop-induced uplift curvature is subdominant.
Above it, a non-sequestered heavy threshold would need to be suppressed by $Q<1$ or computed explicitly.
In the exactly sequestered branch, $c_{\rm mix}=0$, so the mixed one-loop correction vanishes everywhere in the plane.}
\label{fig:one_loop_kahler_exact_scan}
\end{figure}

The interpretation of Fig.~\ref{fig:one_loop_kahler_exact_scan} is direct. 
The horizontal axis is the threshold logarithm $L=\ln(\Lambda^2/M_{\rm Stk}^2)$, and the vertical axis is the uplift fraction $R=V_{\rm up}/V$. 
The color scale gives the exact normalized correction $\Xi=RL/(32\pi^2)$. 
Because $32\pi^2=315.8273408\ldots$, even moderately large threshold logarithms are strongly loop-suppressed. 
The black contour is determined by the exact product condition $RL=3.1582734\ldots$. 
For example, if $R=1$, the threshold logarithm must reach $L_c=3.1583$ before a $Q=1$ non-sequestered threshold produces a percent-level curvature correction. 
If $R=0.5$, the corresponding boundary is pushed to $L_c=6.3165$. 
For $R=0.1$, the boundary lies at $L_c=31.5827$, far outside the plotted range. 
Equivalently, for $L=3$ the critical uplift fraction is $R_c=1.0528$, outside the physical range $R\leq1$, so the correction remains below the percent level throughout the physical region.

Thus the conclusion is stronger than a purely qualitative naturalness statement. 
The one-loop K\"ahler correction factorizes exactly in the sequestered branch and generates no $T$--$\Sigma$ mixing on the inflationary trajectory. 
If non-sequestered thresholds are introduced, their effect is quantitatively controlled by Eq.~\eqref{eq:Delta_eta_loop_estimate}; the normalized correction is the exact function $\Xi=RL/(32\pi^2)$ shown in Fig.~\ref{fig:one_loop_kahler_exact_scan}. 
Therefore the constant St\"uckelberg uplift and the $\alpha$-attractor plateau are stable at one loop unless an unsuppressed messenger sector couples directly to both the inflaton modulus and the St\"uckelberg sector.

\subsection*{Canonical normalization and the E-model plateau}

Along the real trajectory $T=\bar T>0$, the kinetic term of $T$ is
\begin{align}
\mathcal{L}_{\rm kin}
=
K_{T\bar T}\partial T\partial\bar T
=
\frac{3\alpha}{(T+\bar T)^2}\partial T\partial\bar T .
\label{eq:T_kinetic}
\end{align}
Setting
\begin{align}
T=e^{\beta\phi},
\qquad
\beta=\sqrt{\frac{2}{3\alpha}},
\label{eq:canonmap}
\end{align}
one obtains
\begin{align}
\frac{3\alpha}{(T+\bar T)^2}\partial T\partial\bar T
=
\frac{3\alpha}{4T^2}(\partial T)^2
=
\frac{1}{2}(\partial\phi)^2 .
\label{eq:canonical_proof}
\end{align}
Substituting Eq.~\eqref{eq:canonmap} into Eq.~\eqref{eq:VFvalley} gives the E-model potential
\begin{align}
U(\phi)
=
V_0\left(1-e^{-\beta\phi}\right)^2,
\qquad
V_0=\mu^2.
\label{eq:Emodel}
\end{align}
The potential approaches $V_0$ exponentially for $\phi\to+\infty$ and has a supersymmetric minimum at $T=1$ in the absence of the constant uplift.

The derivatives of the tree-level inflaton potential are
\begin{align}
U'(\phi)
=
2\beta V_0 e^{-\beta\phi}
\left(1-e^{-\beta\phi}\right),
\label{eq:Uprime}
\\
U''(\phi)
=
2\beta^2V_0 e^{-\beta\phi}
\left(2e^{-\beta\phi}-1\right).
\label{eq:Udoubleprime}
\end{align}
For the effective single-field potential
\begin{align}
V_{\rm val}(\phi)
=
U(\phi)+V_{\rm up},
\label{eq:Vval_single_field}
\end{align}
the potential slow-roll parameters at fixed field value are therefore
\begin{align}
\epsilon_V(\phi)
=
\frac{1}{2}
\left(
\frac{U'(\phi)}{U(\phi)+V_{\rm up}}
\right)^2
=
2\beta^2
\frac{
V_0^2 e^{-2\beta\phi}
\left(1-e^{-\beta\phi}\right)^2
}{
\left[
V_0\left(1-e^{-\beta\phi}\right)^2+V_{\rm up}
\right]^2
},
\label{eq:epsilon_fixed_phi}
\\
\eta_V(\phi)
=
\frac{U''(\phi)}{U(\phi)+V_{\rm up}}
=
2\beta^2
\frac{
V_0e^{-\beta\phi}
\left(2e^{-\beta\phi}-1\right)
}{
V_0\left(1-e^{-\beta\phi}\right)^2+V_{\rm up}
}.
\label{eq:eta_fixed_phi}
\end{align}
At fixed $\phi$, the uplift suppresses $\epsilon_V$ by increasing the denominator of the potential. At fixed $N_*$, however, the same uplift also shifts the horizon-crossing value $\phi_*$ through the $N$-$\phi$ relation. Consequently, the leading attractor scaling is preserved, while the uplift contributes through finite-$N_*$, logarithmic, and end-point corrections. The detailed $N$-$\phi$ mapping is derived in Section~\ref{sec:background}.

\subsection*{Waterfall sector, angular mode, and effective valley potential}

The hybrid exit is triggered by the charged waterfall fields. We take the $U(1)_X$ D-term potential to be
\begin{align}
V_X
=
\frac{g_X^2}{2}
\left(
|\Psi|^2-|\bar\Psi|^2-v_D^2
\right)^2 .
\label{eq:VX}
\end{align}
Along $\Psi=\bar\Psi=0$, this contributes a constant false-vacuum energy $g_X^2v_D^4/2$. Since this term is independent of $\phi$, it can be absorbed into the effective constant uplift,
\begin{align}
V_{\rm up}^{\rm eff}
=
V_{\rm up}
+
\frac{g_X^2v_D^4}{2}.
\label{eq:Vup_eff}
\end{align}
For notational simplicity, we denote this total constant contribution by $V_{\rm up}$ in the following sections.\footnote{With this convention, $V_{\rm up}$ denotes the total $\phi$-independent positive contribution to the inflationary valley energy. If one keeps the Stückelberg and waterfall false-vacuum pieces separate, one should replace $V_{\rm up}$ everywhere by $V_{\rm up}+g_X^2v_D^4/2$ along the valley.}

The bilinear waterfall contribution from the superpotential gives the coordinate-basis mass coefficient
\begin{align}
m_{F,{\rm coord}}^2(\phi)
=
g^2|1-e^{-\beta\phi}|^2 .
\label{eq:mF_coord}
\end{align}
After canonical normalization of $\Psi$ and $\bar\Psi$, the corresponding physical supersymmetric mass is
\begin{align}
m_{F,{\rm phys}}^2(\phi)
=
(T+\bar T)^{3\alpha}
g^2|1-T^{-1}|^2
=
(2e^{\beta\phi})^{3\alpha}
g^2|1-e^{-\beta\phi}|^2 .
\label{eq:mF_phys}
\end{align}
Similarly, the D-term splitting in canonically normalized variables is
\begin{align}
m_{D,{\rm phys}}^2(\phi)
=
(T+\bar T)^{3\alpha}g_X^2v_D^2 .
\label{eq:mD_phys}
\end{align}
Including the Hubble-induced contribution from Eq.~\eqref{eq:Hubble_mass_minimal}, the waterfall eigenvalues may be written as
\begin{align}
m_\pm^2(\phi)
=
(T+\bar T)^{3\alpha}
\left[
g^2|1-T^{-1}|^2
\pm
g_X^2v_D^2
\right]
+
c_HU(\phi),
\label{eq:waterfall_masses}
\end{align}
where $c_H=1$ for the minimal expansion in Eq.~\eqref{eq:Fterm_waterfall_expansion}, while higher-order K\"ahler corrections can modify this coefficient. Stability of the inflationary valley requires $m_\pm^2(\phi)>0$. The critical point is defined by
\begin{align}
m_-^2(\phi_c)=0.
\label{eq:critical_general}
\end{align}
In the limit where the Hubble-induced term is small compared with the D-term splitting at the waterfall transition, Eq.~\eqref{eq:critical_general} reduces to
\begin{align}
g|1-e^{-\beta\phi_c}|
=
g_Xv_D,
\label{eq:critical_simple}
\end{align}
and therefore
\begin{align}
\phi_c
=
\frac{1}{\beta}
\ln\left(
\frac{1}{1-g_Xv_D/g}
\right),
\qquad
0<\frac{g_Xv_D}{g}<1 .
\label{eq:phic}
\end{align}
The full condition Eq.~\eqref{eq:critical_general} is the appropriate stability criterion in the supergravity effective theory, while Eq.~\eqref{eq:phic} is a useful analytic limit.

Close to the critical point, the lightest waterfall eigenvalue may be expanded as
\begin{align}
m_-^2(\phi)
\simeq
\left.
\frac{d m_-^2}{d\phi}
\right|_{\phi_c}
(\phi-\phi_c).
\label{eq:waterfall_expansion}
\end{align}
The waterfall is prompt when the tachyonic growth rate exceeds the Hubble scale shortly after the critical point,
\begin{align}
\frac{|m_-|}{H}\gtrsim 1 .
\label{eq:prompt_waterfall}
\end{align}
This condition prevents an extended multi-field waterfall phase and justifies treating $\phi_c$ as the effective end point of slow-roll evolution.

The imaginary part of $T$ requires a separate comment. Writing
\begin{align}
T=u e^{i\vartheta},
\qquad
u=e^{\beta\phi},
\label{eq:T_polar}
\end{align}
one obtains, near $\vartheta=0$,
\begin{align}
U(\phi,\vartheta)
=
\mu^2
\left(
1-\frac{2\cos\vartheta}{u}
+\frac{1}{u^2}
\right)
=
U(\phi,0)
+
\mu^2\frac{\vartheta^2}{u}
+
\mathcal{O}(\vartheta^4).
\label{eq:angular_potential}
\end{align}
The angular kinetic term follows from
\begin{align}
K_{T\bar T}\partial T\partial\bar T
\supset
\frac{3\alpha}{4}(\partial\vartheta)^2
=
\frac{1}{2}(\partial\sigma)^2,
\qquad
\sigma=\sqrt{\frac{3\alpha}{2}}\,\vartheta .
\label{eq:angular_canonical}
\end{align}
Thus the canonically normalized angular mass is
\begin{align}
m_\sigma^2
=
\frac{4}{3\alpha}\frac{\mu^2}{u},
\qquad
\frac{m_\sigma^2}{H^2}
\simeq
\frac{4}{\alpha u}
\frac{V_0}{V_0+V_{\rm up}}
\ll 1
\qquad
(u\gg1).
\label{eq:angular_mass}
\end{align}
The angular mode is therefore light on the plateau in the minimal K\"ahler realization. This does not destabilize the real trajectory because $\vartheta=0$ is an extremum and $m_\sigma^2>0$, but it implies that angular fluctuations are generated with amplitude
\begin{align}
\mathcal{P}_{\delta\sigma}
\simeq
\left(
\frac{H_*}{2\pi}
\right)^2 .
\label{eq:angular_fluctuation}
\end{align}
Such fluctuations are harmless only if they are not converted into curvature or matter-isocurvature perturbations. In the trajectory-aligned branch, the turn rate in field space vanishes along $\vartheta=0$, and the waterfall and reheating dynamics are assumed to depend dominantly on the real modulus and charged waterfall fields. If $Y$ denotes a late-time abundance or decay quantity, the induced isocurvature power is parametrically
\begin{align}
\mathcal{P}_{\mathcal S}
\sim
\left(
\frac{\partial\ln Y}{\partial\sigma}
\right)^2
\mathcal{P}_{\delta\sigma}.
\label{eq:isocurvature_estimate}
\end{align}
The assumption
\begin{align}
\left.
\frac{\partial\ln Y}{\partial\sigma}
\right|_{\sigma=0}
\simeq
0
\label{eq:isocurvature_suppression}
\end{align}
therefore suppresses the observable isocurvature mode. Equivalently, one may add a small higher-order K\"ahler correction such as
\begin{align}
\Delta K_T
=
-\kappa_T
\frac{(T-\bar T)^4}{(T+\bar T)^4},
\label{eq:angular_stabilization}
\end{align}
which leaves the real-axis potential unchanged at leading order while increasing $m_\sigma^2/H^2$ for $\kappa_T=\mathcal{O}(1)$. The single-field predictions derived below should be understood as applying to this stabilized or trajectory-aligned branch. If this assumption is relaxed, the system must be treated as a genuine two-field inflationary model.

Combining the F-term plateau, the effective constant uplift, and the waterfall sector, the scalar potential relevant along the inflationary valley is
\begin{align}
V_{\rm val}(\phi)
=
V_0
\left(
1-e^{-\beta\phi}
\right)^2
+
V_{\rm up},
\label{eq:Vtotal_explicit}
\end{align}
where $V_{\rm up}$ denotes the total constant positive contribution according to the convention in Eq.~\eqref{eq:Vup_eff}. This expression is exact at tree level within the specified two-derivative supergravity effective theory and along the truncated valley $S=\Psi=\bar\Psi=0$, subject to the sequestering assumptions discussed above. It provides the effective single-field potential used in the background and perturbation analyses of the following sections.

\section{Background dynamics and the exact $N$-mapping}
\label{sec:background}

\subsection*{Field evolution with constant uplift}

The effective single-field potential along the stabilized or trajectory-aligned inflationary valley is
\begin{align}
V(\phi)
=
U(\phi)+V_{\rm up},
\qquad
U(\phi)
=
V_0\left(1-e^{-\beta\phi}\right)^2,
\qquad
\beta=\sqrt{\frac{2}{3\alpha}} .
\label{eq:V_background}
\end{align}
Here $V_{\rm up}$ denotes the total $\phi$-independent positive contribution along the inflationary valley, including the sequestered St\"uckelberg uplift and, if present, any $\phi$-independent false-vacuum contribution from the waterfall sector. This convention was defined in Sec.~\ref{sec:model}. The constancy of $V_{\rm up}$ relies on the sequestered effective structure discussed there; if Planck-suppressed cross-couplings induce a nontrivial $V_{\rm up}(\phi)$, the formulae below receive additional terms proportional to $dV_{\rm up}/d\phi$ and $d^2V_{\rm up}/d\phi^2$.\footnote{The purpose of this section is to derive the background dynamics in the sequestered branch of the effective theory. The effect of small non-sequestered corrections can be incorporated perturbatively by replacing $V_{\rm up}$ with $V_{\rm up}(\phi)$ and retaining the derivatives of the latter in the slow-roll parameters.}

During slow roll, the scalar equation of motion and Friedmann equation are
\begin{align}
3H\dot\phi\simeq -V_{,\phi},
\qquad
3H^2\simeq V .
\label{eq:slowroll_background_eqs}
\end{align}
Using $N=\ln a$ as the time variable gives
\begin{align}
\frac{d\phi}{dN}
=
\frac{\dot\phi}{H}
\simeq
-\frac{V_{,\phi}}{V}
=
-\frac{U'(\phi)}{U(\phi)+V_{\rm up}} .
\label{eq:dphidN}
\end{align}
This relation is the leading-order slow-roll flow equation. It is not an exact solution of the full second-order background equation, but it is sufficient for deriving the analytic $N$-mapping and leading attractor predictions.

It is useful to introduce
\begin{align}
u
\equiv
e^{\beta\phi},
\qquad
\phi=\frac{1}{\beta}\ln u .
\label{eq:u_def}
\end{align}
In terms of $u$, the potential and its first two derivatives are
\begin{align}
U(u)
=
V_0\left(1-\frac{1}{u}\right)^2
=
V_0\left(1-\frac{2}{u}+\frac{1}{u^2}\right),
\label{eq:U_u}
\\
U'(\phi)
=
2\beta V_0\frac{u-1}{u^2},
\label{eq:Uprime_u}
\\
U''(\phi)
=
2\beta^2V_0\frac{2-u}{u^2}.
\label{eq:Udoubleprime_u}
\end{align}
Since $du/dN=\beta u\,d\phi/dN$, Eq.~\eqref{eq:dphidN} gives
\begin{align}
\frac{du}{dN}
=
-\frac{2\beta^2V_0(u-1)}
{u\left[U(u)+V_{\rm up}\right]} .
\label{eq:du_dN}
\end{align}
The sign is negative because $N=\ln a$ increases with time while the field rolls from larger $u$ toward the critical value $u_c$. Equivalently, the number of $e$-folds remaining until the end of inflation decreases along the trajectory. In what follows, $N(\phi;\phi_c)$ denotes the number of $e$-folds accumulated between the field value $\phi$ and the end point $\phi_c$, so it is positive for $u>u_c$.

\subsection*{Exact analytic $N$--$\phi$ mapping}

The number of $e$-folds between $\phi$ and the end of inflation at $\phi_c$ is
\begin{align}
N(\phi;\phi_c)
=
\int_{t}^{t_c}H\,dt
=
\int_{\phi_c}^{\phi}
\frac{V}{V_{,\phi}}\,d\phi .
\label{eq:N_integral_phi}
\end{align}
Using $d\phi=du/(\beta u)$ and $V_{,\phi}=U'(\phi)$, one obtains
\begin{align}
N(u;u_c)
=
\int_{u_c}^{u}
\frac{U(\tilde u)+V_{\rm up}}
{2\beta^2V_0(\tilde u-1)}
\,d\tilde u .
\label{eq:N_integral_u_start}
\end{align}
The integrand can be simplified as
\begin{align}
\frac{U(u)+V_{\rm up}}{u-1}
&=
\frac{
V_0\left(1-\frac{2}{u}+\frac{1}{u^2}\right)+V_{\rm up}
}{u-1}
\nonumber\\
&=
\frac{
(V_0+V_{\rm up})u^2-2V_0u+V_0
}{
u^2(u-1)
}.
\label{eq:integrand_first}
\end{align}
Equivalently, using
\begin{align}
U(u)
=
V_0\frac{(u-1)^2}{u^2},
\label{eq:U_square_form}
\end{align}
one may write
\begin{align}
\frac{U(u)+V_{\rm up}}{u-1}
=
V_0\frac{u-1}{u^2}
+
V_{\rm up}\frac{1}{u-1}.
\label{eq:integrand_simple}
\end{align}
Therefore
\begin{align}
N(u;u_c)
=
\frac{1}{2\beta^2V_0}
\int_{u_c}^{u}
\left[
V_0\frac{\tilde u-1}{\tilde u^2}
+
V_{\rm up}\frac{1}{\tilde u-1}
\right]
d\tilde u .
\label{eq:N_integral_split_wrong_intermediate}
\end{align}
This form is correct only if the factor $1/u$ from $d\phi=du/(\beta u)$ has been retained consistently. A more direct derivation avoids this possible ambiguity by using
\begin{align}
\frac{V}{V_{,\phi}}\,d\phi
=
\frac{U(u)+V_{\rm up}}
{2\beta V_0(u-1)/u^2}
\frac{du}{\beta u}
=
\frac{u\left[U(u)+V_{\rm up}\right]}
{2\beta^2V_0(u-1)}
\,du .
\label{eq:correct_integrand}
\end{align}
Substituting $U(u)=V_0(u-1)^2/u^2$ gives
\begin{align}
\frac{u\left[U(u)+V_{\rm up}\right]}{u-1}
=
V_0\frac{u-1}{u}
+
V_{\rm up}\frac{u}{u-1}.
\label{eq:correct_integrand_split}
\end{align}
The two terms integrate as
\begin{align}
\int \frac{u-1}{u}\,du
=
u-\ln u,
\qquad
\int \frac{u}{u-1}\,du
=
u+\ln(u-1).
\label{eq:elementary_integrals}
\end{align}
Hence the exact slow-roll $N$--mapping is
\begin{align}
N(u;u_c)
=
\frac{1}{2\beta^2V_0}
\left[
(V_0+V_{\rm up})(u-u_c)
+
V_{\rm up}\ln\frac{u-1}{u_c-1}
-
V_0\ln\frac{u}{u_c}
\right],
\label{eq:Nexact}
\end{align}
where
\begin{align}
u=e^{\beta\phi},
\qquad
u_c=e^{\beta\phi_c}.
\label{eq:u_uc_def}
\end{align}
The leading term is positive for $u>u_c$, as required. The logarithmic terms are subleading on the plateau but are important for finite-$N$ accuracy and for tracking the dependence on the critical point.

For $u\gg1$ and $u\gg u_c$, the leading part of Eq.~\eqref{eq:Nexact} gives
\begin{align}
u_N
\simeq
u_c+
\frac{2\beta^2V_0}{V_0+V_{\rm up}}\,N
=
u_c+
\frac{4}{3\alpha}
\frac{V_0}{V_0+V_{\rm up}}\,N .
\label{eq:Napprox}
\end{align}
Introducing
\begin{align}
f
\equiv
\frac{V_0}{V_0+V_{\rm up}},
\qquad
0<f\leq1,
\label{eq:f_def_background}
\end{align}
this becomes
\begin{align}
u_N
\simeq
u_c+
\frac{4fN}{3\alpha}.
\label{eq:uN_f}
\end{align}
When $V_{\rm up}=0$ and $u_c\ll u_N$, this reduces to the standard E-model attractor relation
\begin{align}
u_N
\simeq
\frac{4N}{3\alpha}.
\label{eq:Nlimit}
\end{align}
For nonzero uplift, Eq.~\eqref{eq:uN_f} shows that the uplift changes the field value at which a fixed number of $e$-folds is generated. This shift is essential when computing observables at fixed $N_*$. At fixed field value, $V_{\rm up}$ suppresses the slow-roll parameter $\epsilon_V$ by increasing the denominator of $V_{,\phi}/V$. At fixed $N_*$, however, the same uplift shifts $u_*$, and the leading $f$-dependence cancels in the attractor limit. Thus the constant uplift does not produce an independent leading $f^2$ suppression of $r$ at fixed $N_*$. Its effect enters through finite-$N_*$ corrections, the logarithmic terms in Eq.~\eqref{eq:Nexact}, and the precise value of the hybrid end point $u_c$.\footnote{This distinction is important because $\epsilon_V(\phi)$ and $\epsilon_V(N)$ are not the same quantity. Statements about suppression at fixed $\phi$ cannot be transferred directly to observables evaluated at fixed $N_*$.}

\subsection*{Slow-roll parameters and leading attractor predictions}

The potential slow-roll parameters are
\begin{align}
\epsilon_V
=
\frac{1}{2}
\left(
\frac{V_{,\phi}}{V}
\right)^2,
\qquad
\eta_V
=
\frac{V_{,\phi\phi}}{V}.
\label{eq:slowroll_def}
\end{align}
Since $V_{,\phi}=U'(\phi)$ and $V_{,\phi\phi}=U''(\phi)$, Eqs.~\eqref{eq:Uprime_u} and \eqref{eq:Udoubleprime_u} give
\begin{align}
\epsilon_V(u)
=
\frac{2\beta^2V_0^2}{\left[U(u)+V_{\rm up}\right]^2}
\frac{(u-1)^2}{u^4},
\label{eq:eps_u_exact}
\\
\eta_V(u)
=
\frac{2\beta^2V_0}{U(u)+V_{\rm up}}
\left(
\frac{2}{u^2}-\frac{1}{u}
\right).
\label{eq:eta_u_exact}
\end{align}
On the plateau, $u\gg1$, these become
\begin{align}
\epsilon_V
\simeq
2\beta^2 f^2\frac{1}{u^2},
\qquad
\eta_V
\simeq
-2\beta^2 f\frac{1}{u}.
\label{eq:eps_eta_plateau_phi}
\end{align}
Substituting the leading $N$--mapping $u_N\simeq 4fN/(3\alpha)$ gives
\begin{align}
\epsilon_V(N)
\simeq
2\beta^2 f^2
\left(
\frac{3\alpha}{4fN}
\right)^2
=
\frac{3\alpha}{4N^2},
\label{eq:epsilon_N_leading}
\\
\eta_V(N)
\simeq
-2\beta^2 f
\left(
\frac{3\alpha}{4fN}
\right)
=
-\frac{1}{N}.
\label{eq:eta_N_leading}
\end{align}
The leading $f$-dependence cancels in both expressions. Therefore the leading-order scalar spectral index and tensor-to-scalar ratio are
\begin{align}
n_s
=
1-6\epsilon_V+2\eta_V
\simeq
1-\frac{2}{N_*}
-\frac{9\alpha}{2N_*^2},
\label{eq:ns_leading_with_eps}
\\
r
=
16\epsilon_V
\simeq
\frac{12\alpha}{N_*^2}.
\label{eq:r_leading}
\end{align}
To leading order in $1/N_*$, this is the usual universal $\alpha$-attractor result,
\begin{align}
n_s
\simeq
1-\frac{2}{N_*},
\qquad
r
\simeq
\frac{12\alpha}{N_*^2}.
\label{eq:nsr_attractor_corrected}
\end{align}
The uplift affects these expressions through subleading terms obtained from the exact mapping Eq.~\eqref{eq:Nexact}. It also affects the overall scalar amplitude because $H_*^2\simeq V_*/3$ depends on the total plateau height. The amplitude normalization therefore fixes a relation among $V_0$, $V_{\rm up}$, $\alpha$, and $N_*$, while the leading tilt and tensor ratio retain the attractor scaling.

The scalar amplitude at leading order is
\begin{align}
A_s
\simeq
\frac{V_*}{24\pi^2\epsilon_{V*}}
=
\frac{
V_0\left(1-u_*^{-1}\right)^2+V_{\rm up}
}{
24\pi^2\epsilon_{V*}
}.
\label{eq:As_background}
\end{align}
On the plateau,
\begin{align}
A_s
\simeq
\frac{V_0+V_{\rm up}}{24\pi^2}
\frac{4N_*^2}{3\alpha}
=
\frac{(V_0+V_{\rm up})N_*^2}{18\pi^2\alpha}.
\label{eq:As_plateau}
\end{align}
Thus, for fixed $A_s$, increasing the constant uplift changes the required value of $V_0$ but does not by itself generate a leading suppression of $r$ at fixed $N_*$.

\subsection*{Finite-$N$ corrections and dependence on the hybrid end point}

The exact relation Eq.~\eqref{eq:Nexact} can be used to compute the finite-$N$ corrections systematically. Let
\begin{align}
u_N
=
\frac{4fN}{3\alpha}
+
u_c
+
\Delta u_{\log},
\label{eq:uN_expansion_form}
\end{align}
where $\Delta u_{\log}$ denotes the correction induced by the logarithmic terms in Eq.~\eqref{eq:Nexact}. Since the logarithms grow only as $\ln u_N$, whereas the leading term grows as $N$, one has
\begin{align}
\frac{\Delta u_{\log}}{u_N}
=
\mathcal{O}
\left(
\frac{\ln N}{N}
\right),
\qquad
\frac{u_c}{u_N}
=
\mathcal{O}
\left(
\frac{u_c}{N}
\right).
\label{eq:finiteN_scaling}
\end{align}
Accordingly,
\begin{align}
\epsilon_V(N)
=
\frac{3\alpha}{4N^2}
\left[
1
+
\delta_\epsilon(f,u_c,N)
\right],
\label{eq:epsilon_delta}
\\
\eta_V(N)
=
-\frac{1}{N}
\left[
1
+
\delta_\eta(f,u_c,N)
\right],
\label{eq:eta_delta}
\end{align}
where $\delta_\epsilon$ and $\delta_\eta$ are controlled by $\ln N/N$ and $u_c/N$ in the plateau regime. The observables can then be written as
\begin{align}
n_s
=
1-\frac{2}{N_*}
+
\delta n_s(f,u_c,N_*),
\label{eq:ns_delta}
\\
r
=
\frac{12\alpha}{N_*^2}
\left[
1+\delta r(f,u_c,N_*)
\right].
\label{eq:r_delta}
\end{align}
The functions $\delta n_s$ and $\delta r$ are not universal because they depend on the exact hybrid end point and on the logarithmic terms in Eq.~\eqref{eq:Nexact}. In the parameter region considered here, these corrections are small compared with the leading attractor terms. This explains why the numerical predictions cluster close to the standard red-tilted attractor line even when $V_{\rm up}\neq0$.

\section{Perturbations and quantum consistency}
\label{sec:perturbations}

\subsection*{Hubble-flow hierarchy and next-to-leading-order spectra}

The perturbations are computed on the stabilized or trajectory-aligned branch described in Sec.~\ref{sec:model}, so that the effective dynamics is single-field and adiabatic. The Hubble-flow parameters are defined by
\begin{align}
\epsilon_1
\equiv
-\frac{\dot H}{H^2},
\qquad
\epsilon_{n+1}
\equiv
\frac{d\ln\epsilon_n}{dN}.
\label{eq:Hubbleflowdef}
\end{align}
At leading order in slow roll,
\begin{align}
\epsilon_1
\simeq
\frac{1}{2}
\left(
\frac{d\phi}{dN}
\right)^2
\simeq
\frac{1}{2}
\left(
\frac{U'(\phi)}{U(\phi)+V_{\rm up}}
\right)^2
=
\epsilon_V .
\label{eq:eps1_slowroll}
\end{align}
Beyond leading order, the Hubble-flow hierarchy should be constructed from the background solution rather than by simply replacing $\epsilon_1$ with $\epsilon_V$ in all expressions. In practice, one may use the exact slow-roll mapping Eq.~\eqref{eq:Nexact} to determine $\phi(N)$ and then compute
\begin{align}
\epsilon_1(N)
=
\frac{1}{2}
\left[
\frac{d\phi(N)}{dN}
\right]^2,
\qquad
\epsilon_2(N)
=
\frac{d\ln\epsilon_1}{dN},
\qquad
\epsilon_3(N)
=
\frac{d\ln\epsilon_2}{dN}.
\label{eq:hubble_hierarchy_from_mapping}
\end{align}
This procedure consistently includes the finite-$N$ and end-point corrections associated with the uplifted hybrid trajectory.

At next-to-leading order in the Stewart--Lyth expansion~\cite{Stewart:1993zq}, the scalar power spectrum evaluated at the pivot scale $k_*$ is
\begin{align}
\mathcal{P}_{\zeta}(k_*)
=
\frac{H_*^2}{8\pi^2\epsilon_{1*}}
\left[
1
-
(2\tilde C+1)\epsilon_{1*}
-
\tilde C\epsilon_{2*}
\right]
+
\mathcal{O}(\epsilon^2),
\label{eq:Ps_NLO}
\end{align}
where
\begin{align}
\tilde C
=
-2+\ln2+\gamma_E
\simeq
-0.7296 .
\label{eq:Ctilde_def}
\end{align}
The spectral index, tensor-to-scalar ratio, and tensor tilt are
\begin{align}
n_s
=
1
-
2\epsilon_{1*}
-
\epsilon_{2*}
-
2\epsilon_{1*}^2
-
(2\tilde C+3)\epsilon_{1*}\epsilon_{2*}
-
\tilde C\epsilon_{2*}\epsilon_{3*}
+
\mathcal{O}(\epsilon^3),
\label{eq:ns_NLO}
\\
r
=
16\epsilon_{1*}
\left[
1+\tilde C(\epsilon_{2*}-2\epsilon_{1*})
\right]
+
\mathcal{O}(\epsilon^3),
\label{eq:r_NLO}
\\
n_t
=
-2\epsilon_{1*}
-
2(1+\tilde C)\epsilon_{1*}\epsilon_{2*}
+
\mathcal{O}(\epsilon^3).
\label{eq:nt_NLO}
\end{align}
These expressions imply the next-to-leading-order single-field consistency relation
\begin{align}
r
=
-8n_t
\left[
1
-
\frac{1+\tilde C}{2}n_t
+
\frac{\tilde C}{2}(n_s-1)
\right]
+
\mathcal{O}(\epsilon^3).
\label{eq:consistency_NLO}
\end{align}
The initial state is chosen to be the Bunch--Davies vacuum,
\begin{align}
v_k
\longrightarrow
\frac{e^{-ik\tau}}{\sqrt{2k}},
\qquad
k|\tau|\to\infty .
\label{eq:BD_vacuum}
\end{align}

In the plateau regime, the leading estimates derived in Sec.~\ref{sec:background} give
\begin{align}
\epsilon_{1*}
\simeq
\frac{3\alpha}{4N_*^2},
\qquad
\epsilon_{2*}
\simeq
\frac{2}{N_*},
\qquad
\epsilon_{3*}
\simeq
\frac{1}{N_*}.
\label{eq:hubble_plateau_estimates}
\end{align}
Substituting these into Eqs.~\eqref{eq:ns_NLO} and \eqref{eq:r_NLO} gives the usual red-tilted attractor behavior with small next-to-leading-order corrections. The constant uplift modifies these values only through the finite-$N_*$ corrections in $\phi(N)$ and through the amplitude normalization. It should therefore be described as producing controlled subleading corrections at fixed $N_*$, not as an independent leading suppression of $r$.

\subsection{Non-Gaussianity}
\label{subsec:nongaussianity}

The same assumptions that justify the single-field power-spectrum calculation also determine the expected size of primordial non-Gaussianity. On the stabilized or trajectory-aligned branch, the curvature perturbation is generated by a single adiabatic degree of freedom during CMB horizon exit. The squeezed-limit bispectrum therefore obeys the standard single-field consistency relation~\cite{Maldacena:2002vr},
\begin{align}
f_{\rm NL}^{\rm local}
=
\frac{5}{12}(1-n_s)
+
\mathcal{O}(\epsilon_i^2),
\label{eq:fNL_local_single_field}
\end{align}
and the equilateral and orthogonal amplitudes are also slow-roll suppressed,
\begin{align}
f_{\rm NL}^{\rm equil},\,f_{\rm NL}^{\rm orth}
=
\mathcal{O}(\epsilon_1,\epsilon_2).
\label{eq:fNL_equil_orth_slowroll}
\end{align}
For the benchmark region with $n_s\simeq0.967$--$0.968$, Eq.~\eqref{eq:fNL_local_single_field} gives $f_{\rm NL}^{\rm local}\simeq1.3\times10^{-2}$, far below current CMB sensitivity~\cite{Planck:2019kim}. The constant uplift changes the background value of $H$ and the exact $N$--$\phi$ mapping, but it does not introduce a new light propagating field or a reduced sound speed in the sequestered branch. Hence it does not generate an additional leading bispectrum source.

The waterfall fields are heavy and stabilized during the observable stage, becoming tachyonic only at the hybrid end point. Consequently, they do not source a large CMB-scale bispectrum in the parameter region used for the predictions above. A larger non-Gaussian signal could arise only outside the effective single-field branch, for example if the angular mode of $T$ or a waterfall direction were light at CMB horizon exit, if the end-of-inflation hypersurface were significantly modulated by a light isocurvature field, or if the trajectory developed a sizable turn rate. These cases require a genuine multi-field analysis and are not included in the single-field predictions quoted here.

\subsection*{Classicality and radiative stability}

The classical slow-roll displacement per Hubble time is
\begin{align}
\Delta\phi_{\rm cl}
\simeq
\left|
\frac{d\phi}{dN}
\right|
=
\sqrt{2\epsilon_1},
\label{eq:classical_displacement}
\end{align}
whereas the quantum fluctuation of a light scalar over one Hubble time is
\begin{align}
\Delta\phi_{\rm q}
\simeq
\frac{H}{2\pi}.
\label{eq:quantum_displacement}
\end{align}
Classical evolution dominates when
\begin{align}
\Delta\phi_{\rm cl}
>
\Delta\phi_{\rm q}.
\label{eq:classical_condition_start}
\end{align}
Using $H^2\simeq V/3$, this condition becomes
\begin{align}
\epsilon_1
>
\frac{V}{24\pi^2}.
\label{eq:classicality_bound}
\end{align}
On the plateau,
\begin{align}
V
\simeq
V_0+V_{\rm up},
\qquad
\epsilon_1
\simeq
\frac{3\alpha}{4N_*^2}.
\label{eq:plateau_classical_quantities}
\end{align}
For $\alpha\lesssim1$ and $N_*=50$--$60$, one obtains $\epsilon_1\sim10^{-4}$--$10^{-3}$, while the observed scalar amplitude requires $V/(24\pi^2)\sim A_s\epsilon_1\sim10^{-13}$--$10^{-12}$. Hence the classicality condition is safely satisfied throughout the observable window. This also excludes the self-reproduction regime for the field values relevant to the CMB.

The one-loop Coleman--Weinberg correction from the waterfall sector has the form
\begin{align}
\Delta V_{\rm CW}^{\rm flat}(\phi)
=
\frac{1}{64\pi^2}
\sum_A
n_A\,m_A^4(\phi)
\left[
\ln\frac{m_A^2(\phi)}{\mu_{\rm R}^2}
-
c_A
\right],
\label{eq:VCW_flat_general}
\end{align}
where $A$ labels the propagating fluctuation eigenstates, $n_A=(-1)^{F_A}N_A$ includes statistics and multiplicity, $c_A$ is the usual scheme-dependent constant, and $\mu_{\rm R}$ is the renormalization scale. 
For the waterfall sector, the relevant masses approach plateau values with exponentially small inflaton dependence,
\begin{align}
m_A^2(\phi)
=
m_{A,\infty}^2
+
\mathcal{O}(e^{-\beta\phi}),
\qquad
\frac{d m_A^2}{d\phi}
=
\mathcal{O}(e^{-\beta\phi}) .
\label{eq:mass_derivative_suppression}
\end{align}
Therefore the constant part of Eq.~\eqref{eq:VCW_flat_general} renormalizes the plateau height, while its contribution to the inflaton slope is suppressed. 
Differentiating Eq.~\eqref{eq:VCW_flat_general} gives
\begin{align}
\frac{d\Delta V_{\rm CW}^{\rm flat}}{d\phi}
=
\frac{1}{32\pi^2}
\sum_A
n_A\,
m_A^2
\frac{d m_A^2}{d\phi}
\left[
\ln\frac{m_A^2}{\mu_{\rm R}^2}
-
c_A
+
\frac{1}{2}
\right],
\label{eq:VCW_flat_derivative}
\end{align}
which is of order $\mathcal{O}(e^{-\beta\phi})$ on the plateau. 
Choosing $\mu_{\rm R}$ close to the characteristic heavy mass scale minimizes large logarithms.

The sequestered St\"uckelberg sector can be included in the same expression. 
Let $M_D$ denote the mass scale of the heavy St\"uckelberg multiplet and the corresponding $U(1)_D$ vector multiplet after the St\"uckelberg mechanism. 
In the exactly sequestered branch,
\begin{align}
K
=
K_{\rm infl}(T,S,\Psi,\bar\Psi)
+
K_{\rm Stk}(Y),
\qquad
Y=\Sigma+\bar\Sigma+q_DV_D,
\qquad
W_{\rm tot}=W_{\rm infl},
\qquad
\partial_\Sigma W_{\rm tot}=0,
\label{eq:sequestered_inputs_CW}
\end{align}
the heavy St\"uckelberg-sector eigenvalues are independent of the inflaton at tree level:
\begin{align}
M_{D,a}^2(\phi)
=
M_{D,a}^2
\qquad
\Longrightarrow
\qquad
\frac{dM_{D,a}^2}{d\phi}=0 .
\label{eq:Stk_mass_factorized}
\end{align}
Hence its flat-space Coleman--Weinberg contribution,
\begin{align}
\Delta V_{\rm CW}^{D}
=
\frac{1}{64\pi^2}
\sum_{a\in D}
n_a\,M_{D,a}^4
\left[
\ln\frac{M_{D,a}^2}{\mu_{\rm R}^2}
-
c_a
\right],
\label{eq:VCW_Stk_constant}
\end{align}
is a constant threshold correction. 
It can be absorbed into the renormalized value of the constant uplift and does not generate an inflaton force. 
Thus, in the factorized branch, the sequestered uplift sector does not add a new Coleman--Weinberg slope.

If the sequestering is mildly violated, the heavy St\"uckelberg thresholds may acquire weak inflaton dependence. 
The leading dependence has the same origin as the one-loop K\"ahler mixing discussed in Sec.~II:
\begin{align}
M_{D,a}^2(\phi)
=
M_{D,a}^2
\left[
1+
\epsilon_{\Sigma,a}^{(1)}
\widehat{\mathcal F}_{\Sigma,a}^{(1)}(\phi)
+
\mathcal{O}\!\left((\epsilon_{\Sigma,a}^{(1)})^2\right)
\right],
\label{eq:Stk_mass_nonseq}
\end{align}
with
\begin{align}
\epsilon_{\Sigma,a}^{(1)}
\sim
\frac{c_{{\rm mix},a}}{32\pi^2}
\ln\left(\frac{\Lambda^2}{M_{D,a}^2}\right).
\label{eq:epsilon_loop_CW}
\end{align}
Substituting Eq.~\eqref{eq:Stk_mass_nonseq} into Eq.~\eqref{eq:VCW_flat_derivative} shows that any St\"uckelberg-induced Coleman--Weinberg slope is controlled by the same non-sequestered threshold parameter $c_{\rm mix}$:
\begin{align}
\frac{d\Delta V_{\rm CW}^{D}}{d\phi}
\simeq
\frac{1}{32\pi^2}
\sum_{a\in D}
n_a\,M_{D,a}^4\,
\epsilon_{\Sigma,a}^{(1)}
\frac{d\widehat{\mathcal F}_{\Sigma,a}^{(1)}}{d\phi}
\left[
\ln\frac{M_{D,a}^2}{\mu_{\rm R}^2}
-
c_a
+
\frac{1}{2}
\right].
\label{eq:VCW_Stk_slope_nonseq}
\end{align}
Therefore the St\"uckelberg sector produces no inflaton-dependent Coleman--Weinberg contribution in the exactly sequestered branch, while nonzero contributions require precisely the type of non-sequestered heavy threshold already bounded by Eq.~\eqref{eq:Delta_eta_loop_estimate}. 
This is the Coleman--Weinberg counterpart of the one-loop K\"ahler estimate.

We now justify the use of the flat-space Coleman--Weinberg form during inflation. 
Since the inflationary background is quasi-de Sitter, the one-loop potential should in principle be computed in curved spacetime. 
The curved-spacetime effective potential is known to receive curvature-dependent contributions through curvature-corrected masses, curvature-dependent running scales, and purely gravitational counterterms~\cite{Markkanen:2018bfx}.
For a scalar fluctuation, the ultraviolet heat-kernel form replaces the flat-space mass by
\begin{align}
m_A^2(\phi)
\quad\longrightarrow\quad
\mathcal{M}_A^2(\phi,R_4)
=
m_A^2(\phi)
+
\zeta_A R_4,
\label{eq:curved_mass_shift}
\end{align}
where $R_4$ is the four-dimensional Ricci scalar and $\zeta_A$ is an order-one coefficient determined by the spin and nonminimal curvature coupling of the fluctuation. 
For a minimally coupled scalar, $\zeta_A=-1/6$ in the standard heat-kernel convention. 
In quasi-de Sitter space,
\begin{align}
R_4
=
12H^2
+
\mathcal{O}(\epsilon H^2)
\simeq
4V_{\rm val}(\phi),
\label{eq:dS_curvature}
\end{align}
in reduced Planck units. 
The corresponding curved-spacetime Coleman--Weinberg correction can be written schematically as
\begin{align}
\Delta V_{\rm CW}^{\rm dS}(\phi)
=
\frac{1}{64\pi^2}
\sum_A
n_A\,
\mathcal{M}_A^4(\phi,R_4)
\left[
\ln\frac{\mathcal{M}_A^2(\phi,R_4)}{\mu_{\rm R}^2}
-
c_A
\right]
+
\Delta V_{\rm grav}(R_4,R_{\mu\nu}R^{\mu\nu},\ldots),
\label{eq:VCW_dS_general}
\end{align}
where $\Delta V_{\rm grav}$ denotes the curvature-only counterterms and higher-curvature operators generated by renormalization.

The flat-space approximation is justified when the heavy fields integrated out in the Coleman--Weinberg potential satisfy
\begin{align}
m_A^2(\phi)
\gg
R_4
\simeq
12H^2 .
\label{eq:heavy_curvature_hierarchy}
\end{align}
In that regime,
\begin{align}
\mathcal{M}_A^4
\ln\frac{\mathcal{M}_A^2}{\mu_{\rm R}^2}
=
m_A^4
\ln\frac{m_A^2}{\mu_{\rm R}^2}
+
2\zeta_A R_4 m_A^2
\left[
\ln\frac{m_A^2}{\mu_{\rm R}^2}
+
\frac{1}{2}
\right]
+
\mathcal{O}(R_4^2).
\label{eq:curved_CW_expansion}
\end{align}
The first term is the flat-space Coleman--Weinberg contribution in Eq.~\eqref{eq:VCW_flat_general}. 
The term linear in $R_4$ renormalizes the gravitational coupling and nonminimal curvature operators. 
After the gravitational couplings are renormalized, the remaining curvature-dependent contribution to the scalar potential is of order
\begin{align}
\Delta V_{\rm CW}^{\rm curv}
\sim
\frac{1}{64\pi^2}
\sum_A
n_A\,\mathcal{O}(R_4^2)
\sim
\mathcal{O}\!\left(\frac{H^4}{16\pi^2}\right).
\label{eq:curvature_residual_size}
\end{align}
Compared with the inflationary energy density $V\simeq3H^2$, this gives
\begin{align}
\frac{\Delta V_{\rm CW}^{\rm curv}}{V}
\sim
\mathcal{O}\!\left(\frac{H^2}{16\pi^2}\right).
\label{eq:curvature_energy_ratio}
\end{align}
Using the observed scalar amplitude,
\begin{align}
A_s
=
\frac{H_*^2}{8\pi^2\epsilon_{1*}},
\label{eq:As_H_relation}
\end{align}
one obtains
\begin{align}
\frac{H_*^2}{16\pi^2}
=
\frac{A_s\epsilon_{1*}}{2}.
\label{eq:curvature_As_suppression}
\end{align}
For $\epsilon_{1*}\sim10^{-4}$--$10^{-3}$ and $A_s\simeq2.1\times10^{-9}$, the residual curvature contribution is of order
\begin{align}
\frac{\Delta V_{\rm CW}^{\rm curv}}{V}
\lesssim
10^{-12},
\label{eq:curvature_negligible_number}
\end{align}
up to order-one multiplicity and curvature-coupling factors. 
Its contribution to the inflaton slope is correspondingly negligible compared with the tree-level slow-roll slope.

There is also a possible curvature correction to the logarithmic scale choice. 
The optimal curved-spacetime renormalization scale is set by
\begin{align}
\mu_{\rm R}^2
\sim
m_A^2+\zeta_A R_4 .
\label{eq:curved_scale_choice}
\end{align}
For the heavy waterfall and St\"uckelberg thresholds satisfying Eq.~\eqref{eq:heavy_curvature_hierarchy}, this differs from the flat-space choice $\mu_{\rm R}^2\sim m_A^2$ only by a relative correction
\begin{align}
\frac{\delta\mu_{\rm R}^2}{\mu_{\rm R}^2}
\sim
\frac{\zeta_A R_4}{m_A^2}
\ll
1 .
\label{eq:scale_choice_suppression}
\end{align}
Thus curvature-induced running does not modify the Coleman--Weinberg slope at the accuracy relevant for the inflationary observables.

The assumptions behind the radiative-stability statement are therefore explicit. 
First, the waterfall and St\"uckelberg multiplets that are integrated out in the Coleman--Weinberg potential are heavy compared with the Hubble scale during the observable stage,
\begin{align}
m_A^2/H^2\gg1 .
\label{eq:heavy_field_condition_CW}
\end{align}
Second, the St\"uckelberg sector is sequestered so that its heavy masses are $\phi$-independent, or else any residual $\phi$-dependence is controlled by the loop-suppressed mixing coefficient in Eq.~\eqref{eq:epsilon_loop_CW}. 
Third, curvature-only terms generated by the de Sitter background are absorbed into the renormalized gravitational sector, with residual effects suppressed by $H^2/(16\pi^2)$ relative to the inflationary energy density. 
Under these conditions, Eq.~\eqref{eq:VCW_flat_general} is sufficient for the radiative-stability estimate used here, and the sequestered St\"uckelberg uplift does not destabilize the $\alpha$-attractor plateau.
\section{Post-inflationary dynamics and vacuum structure}
\label{sec:postinflation}

\subsection*{Reheating and $N_*$--$k_*$ mapping}

The observable predictions of an inflationary model are evaluated at the time when a comoving pivot scale $k_*$ exits the Hubble radius. The corresponding number of $e$-folds $N_*$ depends on the inflationary scale, the end of inflation, and the subsequent reheating history. Assuming entropy conservation after reheating, the standard matching relation gives~\cite{Liddle:2003as,Dai:2014jja}
\begin{align}
N_*
\simeq
57
-\ln\frac{k_*}{0.05\,{\rm Mpc}^{-1}}
+\frac{1}{4}\ln\frac{V_*}{(10^{16}\,{\rm GeV})^4}
+\frac{1}{4}\ln\frac{V_*}{V_{\rm end}}
+\frac{1-3w_{\rm reh}}{12(1+w_{\rm reh})}
\ln\frac{\rho_{\rm reh}}{\rho_{\rm end}} .
\label{eq:Nstar_kstar}
\end{align}
Here $V_*=V(\phi_*)$ is the potential energy at horizon crossing, $V_{\rm end}=V(\phi_{\rm end})$ is the energy density at the end of inflation, $\rho_{\rm reh}$ is the energy density at the end of reheating, and $w_{\rm reh}$ is the effective equation-of-state parameter during reheating. In the hybrid scenario considered here, the end of inflation is controlled by the waterfall instability rather than by the condition $\epsilon_V=1$. Thus,
\begin{align}
\phi_{\rm end}=\phi_c,
\qquad
m_-^2(\phi_c)=0,
\label{eq:end_condition_postinflation}
\end{align}
and the energy density at the end of the slow-roll phase is
\begin{align}
V_{\rm end}
=
U(\phi_c)+V_{\rm up}.
\label{eq:Vend_postinflation}
\end{align}
The notation $V_{\rm up}$ denotes the effective constant contribution along the inflationary valley, including the sequestered Stückelberg uplift and any $\phi$-independent false-vacuum contribution that has been absorbed into it according to the convention introduced in Sec.~\ref{sec:model}.\footnote{If the Stückelberg uplift and the false-vacuum energy of the $U(1)_X$ waterfall sector are kept separate, one should replace $V_{\rm up}$ in the valley potential by $V_{\rm up}+g_X^2v_D^4/2$. The single symbol $V_{\rm up}$ is used here only as a shorthand for the total constant contribution during inflation.}

The matching formula in Eq.~\eqref{eq:Nstar_kstar} shows that the uncertainty in reheating translates into an uncertainty in $N_*$. For a fixed pivot scale, the reheating contribution is controlled by
\begin{align}
\Delta N_{\rm reh}
=
\frac{1-3w_{\rm reh}}{12(1+w_{\rm reh})}
\ln\frac{\rho_{\rm reh}}{\rho_{\rm end}} .
\label{eq:DeltaN_reh}
\end{align}
For instantaneous reheating, $\rho_{\rm reh}\simeq\rho_{\rm end}$ and $\Delta N_{\rm reh}\simeq0$. For a matter-like reheating phase with $w_{\rm reh}=0$, one obtains
\begin{align}
\Delta N_{\rm reh}
=
\frac{1}{12}
\ln\frac{\rho_{\rm reh}}{\rho_{\rm end}},
\label{eq:DeltaN_matter_reh}
\end{align}
which is negative because $\rho_{\rm reh}<\rho_{\rm end}$. Thus a prolonged matter-like reheating phase lowers the preferred value of $N_*$. For radiation-like reheating with $w_{\rm reh}=1/3$, the last term in Eq.~\eqref{eq:Nstar_kstar} vanishes.

The energy stored in the inflaton and waterfall sector after the instability is released into scalar oscillations and subsequently into radiation. If perturbative decays dominate, and if the total decay width into light degrees of freedom is $\Gamma$, the reheating temperature is
\begin{align}
T_{\rm RH}
\simeq
\left(
\frac{90}{\pi^2 g_*}
\right)^{1/4}
\sqrt{\Gamma M_{\rm Pl}},
\label{eq:TRH_general}
\end{align}
where $g_*$ is the number of relativistic degrees of freedom at reheating. For a Yukawa portal with coupling $y$ and a characteristic oscillation mass $m$, one has
\begin{align}
\Gamma
\simeq
\frac{y^2}{8\pi}m ,
\qquad
T_{\rm RH}
\simeq
\left(
\frac{90}{\pi^2 g_*}
\right)^{1/4}
\sqrt{\frac{y^2mM_{\rm Pl}}{8\pi}} .
\label{eq:TRH_yukawa}
\end{align}
In a hybrid model, nonperturbative effects such as tachyonic preheating and parametric resonance may also occur immediately after the waterfall transition~\cite{Felder:2001kt}. These processes can transfer a significant fraction of the energy into waterfall fluctuations before the system thermalizes. For the purpose of the CMB-scale predictions, this uncertainty is efficiently parametrized by $w_{\rm reh}$ and $\rho_{\rm reh}$ in Eq.~\eqref{eq:Nstar_kstar}.

The dependence of the observables on the reheating history is mild because the leading attractor behavior is controlled by $N_*$. To leading order,
\begin{align}
n_s
\simeq
1-\frac{2}{N_*},
\qquad
r
\simeq
\frac{12\alpha}{N_*^2},
\label{eq:reheating_dependence_nsr}
\end{align}
up to finite-$N_*$, uplift, and hybrid-end corrections. A small shift $\Delta N_*$ therefore gives
\begin{align}
\Delta n_s
\simeq
\frac{2}{N_*^2}\Delta N_*,
\qquad
\frac{\Delta r}{r}
\simeq
-\frac{2\Delta N_*}{N_*}.
\label{eq:Delta_nsr_reheating}
\end{align}
For $N_*=50$--$60$, a reheating uncertainty of a few $e$-folds shifts $n_s$ at the level of $10^{-3}$ and changes $r$ by a relative amount of order $10\%$. This is relevant for precision comparisons, but it does not change the qualitative red-tilted $\alpha$-attractor character of the model.

\subsection*{Stabilization, defects, and vacuum energy after the waterfall}

The scalar component of the stabilizer field $S$ must remain heavy during inflation so that it does not participate in the slow-roll dynamics. A standard way to achieve this is to add a quartic correction to the K\"ahler potential,
\begin{align}
\Delta K_S
=
-\kappa_S
(T+\bar T)^{-3\alpha}|S|^4,
\qquad
\kappa_S>0 .
\label{eq:KS_quartic}
\end{align}
Expanding the F-term potential around $S=0$ gives a stabilizer mass of the form
\begin{align}
m_S^2
=
c_SH^2
+
\mathcal{O}(e^{-\beta\phi}),
\label{eq:mS_general}
\end{align}
where $c_S$ is positive and depends on $\kappa_S$ and on the detailed normalization of the higher-order operator. For $\kappa_S=\mathcal{O}(1)$, one obtains $m_S^2\gtrsim H^2$, which stabilizes $S$ at the origin throughout the observable stage of inflation. Alternatively, one may impose the nilpotency constraint
\begin{align}
S^2=0,
\label{eq:nilpotent_S}
\end{align}
which removes the scalar component of $S$ from the physical spectrum while preserving its F-term contribution~\cite{Ferrara:2014kva}. In either description, the role of $S$ is to generate the inflaton potential without introducing an additional light scalar degree of freedom.

The angular component of $T$ requires a separate assumption, as discussed in Sec.~\ref{sec:model}. In the minimal K\"ahler realization, ${\rm Im}\,T$ can be lighter than $H$ on the plateau. The single-field treatment remains valid if the trajectory is aligned along ${\rm Im}\,T=0$ with negligible turn rate and if the angular fluctuation is not converted into a late-time isocurvature perturbation. Equivalently, a higher-order angular-stabilizing K\"ahler correction may be included so that $m_{{\rm Im}\,T}^2/H^2\gtrsim1$ during the observable stage. This assumption is part of the effective single-field branch analyzed in the following observational comparison.

Since the waterfall fields are charged under $U(1)_X$, the waterfall transition breaks this gauge symmetry. If the vacuum manifold has a nontrivial first homotopy group, cosmic strings can form. For a local Abelian string, the tension is approximately
\begin{align}
\mu_{\rm str}
\simeq
2\pi v_D^2\,B(\beta_{\rm AH}),
\label{eq:string_tension_general}
\end{align}
where $B(\beta_{\rm AH})$ is an order-one function of the Abelian-Higgs parameter $\beta_{\rm AH}$. In the BPS-like limit, this is often estimated as
\begin{align}
\mu_{\rm str}
\simeq
2\pi v_D^2 .
\label{eq:string_tension_bps}
\end{align}
Restoring the reduced Planck mass, the dimensionless string tension is
\begin{align}
G\mu_{\rm str}
\simeq
\frac{\mu_{\rm str}}{8\pi M_{\rm Pl}^2}
\simeq
\frac{v_D^2}{4M_{\rm Pl}^2}
\end{align}
up to order-one convention-dependent factors. In reduced Planck units, the observational requirement $G\mu_{\rm str}\lesssim10^{-7}$--$10^{-8}$ implies roughly
\begin{align}
v_D
\lesssim
10^{-3}\text{--}10^{-4},
\label{eq:vD_bound}
\end{align}
depending on the string network assumptions and on the precise normalization of the tension~\cite{Planck:2018jri,Blanco-Pillado:2013qja}. This constraint can be satisfied by taking the waterfall scale below the inflationary plateau scale or by embedding the symmetry-breaking sector in a way that avoids stable string formation. The hidden $U(1)_D$ Stückelberg sector is sequestered and does not break through the waterfall fields, so it does not generate the same class of cosmic strings.

The vacuum-energy accounting must be defined carefully. With the positive-definite waterfall D-term
\begin{align}
V_X
=
\frac{g_X^2}{2}
\left(
|\Psi|^2-|\bar\Psi|^2-v_D^2
\right)^2 ,
\label{eq:VX_post}
\end{align}
the energy at $\Psi=\bar\Psi=0$ is
\begin{align}
V_X\big|_{\Psi=\bar\Psi=0}
=
\frac{g_X^2v_D^4}{2}.
\label{eq:VX_false_vacuum}
\end{align}
This contribution is independent of $\phi$ and was absorbed into the effective constant $V_{\rm up}$ along the inflationary valley. After the waterfall, the charged fields relax to a symmetry-breaking minimum satisfying
\begin{align}
|\Psi|^2-|\bar\Psi|^2=v_D^2,
\label{eq:Dflat_waterfall_vacuum}
\end{align}
so that
\begin{align}
V_X\big|_{\rm broken}=0 .
\label{eq:VX_broken_zero}
\end{align}
The F-term contribution from the inflaton-stabilizer sector also vanishes at $T=1$ and $S=0$,
\begin{align}
U(T=1)=0.
\label{eq:U_vac_zero}
\end{align}
Therefore the post-waterfall vacuum energy is controlled by the residual hidden-sector offset and by any additional late-time tuning sector,
\begin{align}
V_{\rm vac}
=
V_{\rm hid}^{\rm res}
+
V_{\rm tune}.
\label{eq:Vvac_general}
\end{align}
A phenomenologically acceptable late-time vacuum requires
\begin{align}
V_{\rm vac}
\simeq
0
\label{eq:Minkowski_tuning}
\end{align}
on inflationary scales, or a small positive value corresponding to the observed dark-energy density. This tuning is independent of the slow-roll observables, provided the tuning sector is sequestered from the inflaton during the observable stage. Thus the inflationary predictions depend on the valley potential $U(\phi)+V_{\rm up}$, whereas the late-time cosmological constant is fixed by the residual vacuum-energy tuning after the waterfall.\footnote{The subtraction of the late-time cosmological constant should not be confused with a negative contribution from the positive-definite $U(1)_X$ D-term. The $U(1)_X$ sector contributes a false-vacuum energy during inflation and relaxes to zero in the broken phase. Any remaining cancellation is attributed to the hidden or tuning sector.}

\section{Comparison with observations and parameter dependence}
\label{sec:obs}

\begin{figure}[t]
  \centering
  \includegraphics[width=\textwidth]{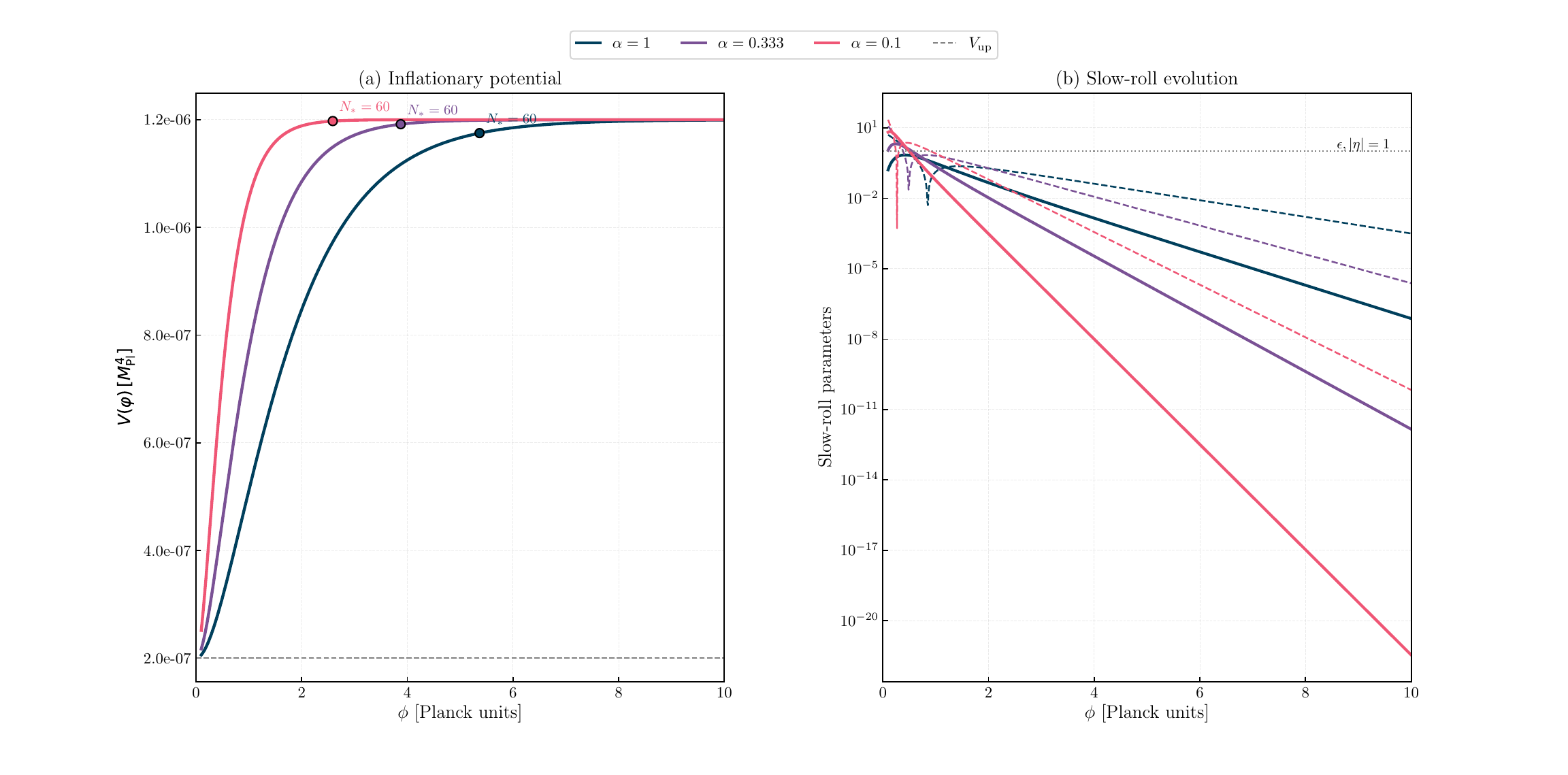}
  \caption{Hybrid $\alpha$-attractor potential and slow-roll parameters for values of $\alpha$. The left panel shows the valley potential $V(\varphi)=V_0(1-e^{-\beta\varphi})^2+V_{\rm up}$, with $\beta=\sqrt{2/(3\alpha)}$. The left vertical axis in the figure itself is labeled $V(\varphi)\,[M_{\rm Pl}^{4}]$, i.e. the scalar potential in reduced Planck units with $M_{\rm Pl}=1$. The right panel shows the corresponding slow-roll parameters $\epsilon_V(\varphi)$ and $|\eta_V(\varphi)|$. Smaller $\alpha$ gives a flatter approach to the plateau and lowers the tensor amplitude through the usual attractor scaling. The constant uplift changes the total energy density and the exact $N$--$\varphi$ relation, while leaving the leading red-tilted attractor behavior intact at fixed $N_*$.}
  \label{fig:Vphi_slowroll}
\end{figure}

Figure~\ref{fig:Vphi_slowroll} displays the inflationary potential and the slow-roll parameters for values $\alpha=1.0$, $\alpha=1/3$, and $\alpha=0.1$. Along the real valley, the potential is
\begin{align}
V(\phi)
=
V_0
\left(
1-e^{-\beta\phi}
\right)^2
+
V_{\rm up},
\qquad
\beta=\sqrt{\frac{2}{3\alpha}} .
\label{eq:V_obs_section}
\end{align}
For large $\phi$, the potential approaches a plateau with height $V_0+V_{\rm up}$. The slope and curvature are controlled by $U'(\phi)$ and $U''(\phi)$, since the uplift is constant in the sequestered branch. The slow-roll parameters are therefore
\begin{align}
\epsilon_V(\phi)
=
\frac{1}{2}
\left(
\frac{U'(\phi)}{U(\phi)+V_{\rm up}}
\right)^2,
\qquad
\eta_V(\phi)
=
\frac{U''(\phi)}{U(\phi)+V_{\rm up}} .
\label{eq:eps_eta_obs_section}
\end{align}
Both quantities decrease along the plateau. The observable values are obtained after using the exact $N$--$\phi$ mapping derived in Sec.~\ref{sec:background}, rather than by evaluating the fixed-field expressions alone. This distinction is important because the uplift suppresses $\epsilon_V$ at fixed $\phi$, while simultaneously shifting the field value $\phi_*$ associated with a fixed $N_*$.

The predictions in the $(n_s,r)$ plane are controlled primarily by the attractor relation
\begin{align}
n_s
\simeq
1-\frac{2}{N_*},
\qquad
r
\simeq
\frac{12\alpha}{N_*^2}
\left[
1+\delta r(f,u_c,N_*)
\right],
\label{eq:nsr_obs_corrected}
\end{align}
where
\begin{align}
f
\equiv
\frac{V_0}{V_0+V_{\rm up}},
\qquad
0<f\leq1,
\label{eq:f_obs_def}
\end{align}
and $\delta r(f,u_c,N_*)$ denotes the subleading correction generated by the logarithmic terms in the exact $N$--mapping and by the hybrid end point $u_c=e^{\beta\phi_c}$. The leading $f^2$ suppression that appears in $\epsilon_V$ at fixed field value is largely compensated at fixed $N_*$ by the uplift-dependent shift of $\phi_*$. Consequently, the numerical dependence of $r$ on $f$ is mild, as shown in Table~\ref{tab:nsr_points}. The dominant parameter controlling the vertical position of the model in the $(n_s,r)$ plane is $\alpha$.

The scalar tilt remains close to the standard red-tilted attractor value. For $N_*=50$, one obtains $n_s\simeq0.960$--$0.962$, while for $N_*=60$ one obtains $n_s\simeq0.967$--$0.968$. These values are characteristic of Planck-compatible $\alpha$-attractor models. They are also compatible with the broader parameter space allowed by recent ACT DR6 and DESI DR2 analyses, but they should not be interpreted as targeting the central ACT-preferred value near $n_s\simeq0.974$. The model therefore remains a standard red-tilted attractor construction rather than a mechanism designed to explain a possible higher-$n_s$ hint.

The tensor amplitude decreases with decreasing $\alpha$. For $\alpha=1$ and $N_*=60$, Table~\ref{tab:nsr_points} gives $r\simeq3\times10^{-3}$, while for $\alpha=0.1$ it gives $r\simeq3\times10^{-4}$. These values are below current upper bounds and can lie within the target range of future $B$-mode searches depending on the experimental sensitivity and foreground control. The weak dependence on $f$ in the table reflects the compensation between the fixed-field suppression of $\epsilon_V$ and the shifted horizon-crossing point at fixed $N_*$.

\begin{figure}[t]
  \centering
  \includegraphics[width=\textwidth]{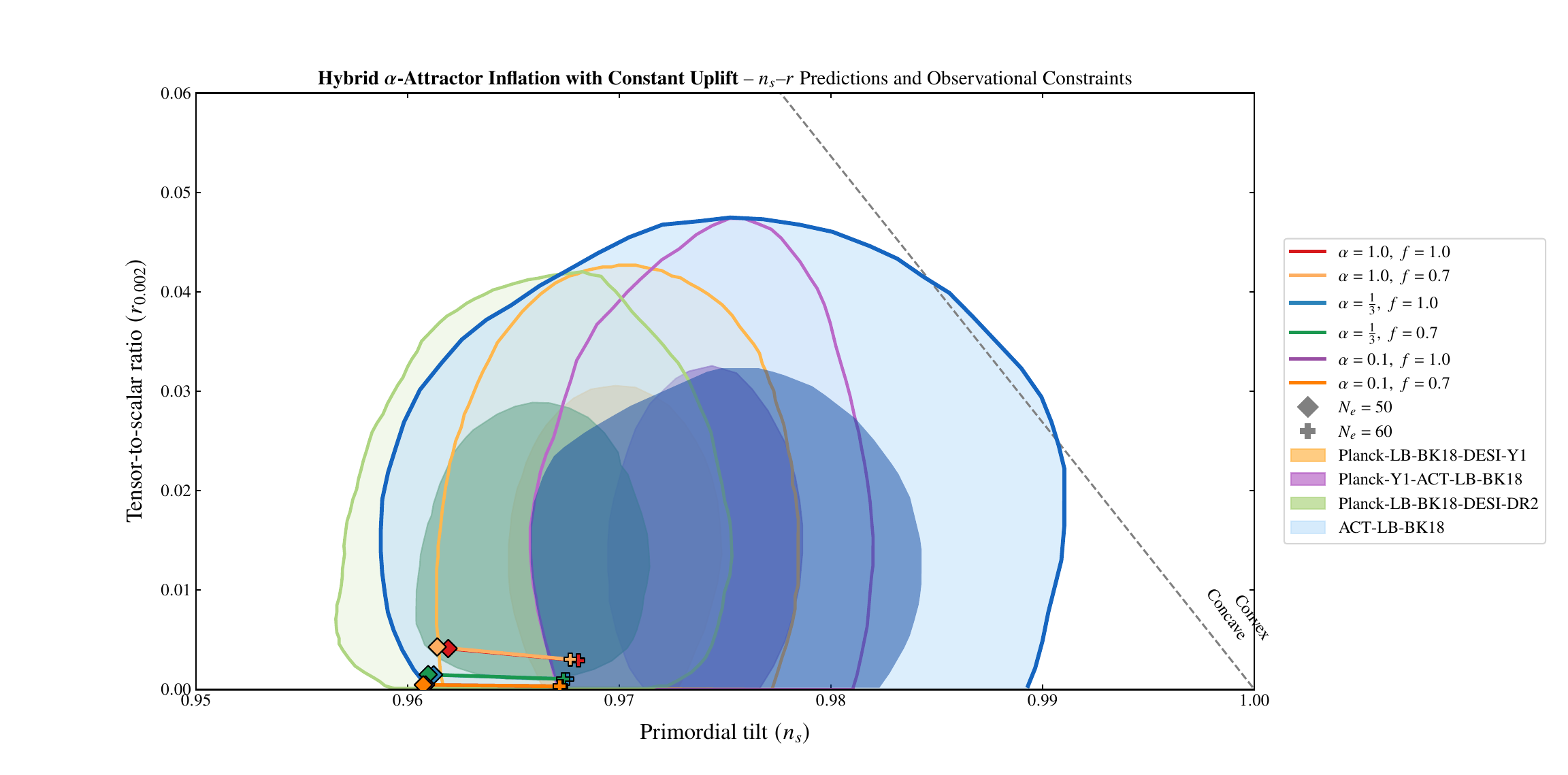}
  \caption{Predictions of the hybrid $\alpha$-attractor model in the $(n_s,r)$ plane. The points correspond to the benchmark values listed in Table~\ref{tab:nsr_points} for $N_*=50$ and $N_*=60$. Decreasing $\alpha$ lowers $r$ through the usual attractor scaling, while changing $f=V_0/(V_0+V_{\rm up})$ produces only a mild shift once the exact $N$--$\phi$ mapping and the hybrid end point are included. The model remains in the standard red-tilted $\alpha$-attractor region and is compatible with current CMB constraints without being designed to reproduce the central value of the ACT-preferred higher-$n_s$ region.}
  \label{fig:nsr}
\end{figure}

\begin{table}[t]
\centering
\setlength{\tabcolsep}{10pt}
\renewcommand{\arraystretch}{1.2}
\begin{tabular}{lcc}
\hline
Model & $N_*$ & $(n_s,\;r)$ \\
\hline
Hybrid $\alpha$-attractor $(\alpha=1.0,\; f=1.0)$ & $50$ & $(0.961913,\;0.004114)$ \\
                                                   & $60$ & $(0.968070,\;0.002918)$ \\
Hybrid $\alpha$-attractor $(\alpha=1.0,\; f=0.7)$ & $50$ & $(0.961396,\;0.004274)$ \\
                                                   & $60$ & $(0.967679,\;0.003018)$ \\
Hybrid $\alpha$-attractor $(\alpha=1/3,\; f=1.0)$ & $50$ & $(0.961228,\;0.001474)$ \\
                                                   & $60$ & $(0.967547,\;0.001036)$ \\
Hybrid $\alpha$-attractor $(\alpha=1/3,\; f=0.7)$ & $50$ & $(0.960966,\;0.001500)$ \\
                                                   & $60$ & $(0.967353,\;0.001052)$ \\
Hybrid $\alpha$-attractor $(\alpha=0.1,\; f=1.0)$ & $50$ & $(0.960847,\;0.000457)$ \\
                                                   & $60$ & $(0.967263,\;0.000320)$ \\
Hybrid $\alpha$-attractor $(\alpha=0.1,\; f=0.7)$ & $50$ & $(0.960737,\;0.000460)$ \\
                                                   & $60$ & $(0.967184,\;0.000322)$ \\
\hline
\end{tabular}
\caption{Numerical benchmark predictions in the $(n_s,r)$ plane. The results show the standard red-tilted attractor behavior, with $n_s$ governed mainly by $N_*$ and $r$ governed mainly by $\alpha$. The dependence on $f=V_0/(V_0+V_{\rm up})$ is mild because the fixed-field suppression of $\epsilon_V$ is compensated at fixed $N_*$ by the uplift-dependent shift in the horizon-crossing field value.}
\label{tab:nsr_points}
\end{table}

\section{Conclusion}
\label{sec:conclusion}

We have constructed a four-dimensional $\mathcal{N}=1$ supergravity effective realization of hybrid inflation within the framework of $\alpha$-attractors. The model combines an inflaton modulus $T$, a stabilizer field $S$, and a pair of oppositely charged waterfall multiplets $(\Psi,\bar\Psi)$ with a sequestered Stückelberg $U(1)_D$ sector. Along the real inflationary valley, the stabilizer F-term generates the E-model plateau
\[
U(\phi)
=
V_0
\left(
1-e^{-\beta\phi}
\right)^2,
\qquad
\beta
=
\sqrt{\frac{2}{3\alpha}},
\]
while the hidden Stückelberg sector contributes an approximately constant positive uplift $V_{\rm up}$. The construction should be understood as a controlled supergravity embedding rather than a UV-complete theory by itself. Its consistency relies on the sequestered effective structure discussed above, in which Planck-suppressed cross-couplings between the inflaton and uplift sectors are absent or sufficiently suppressed.

The analytic control of the model follows from the closed-form slow-roll $N$--$\phi$ mapping. Defining
\[
u=e^{\beta\phi},
\qquad
u_c=e^{\beta\phi_c},
\]
we obtained
\[
N(u;u_c)
=
\frac{1}{2\beta^2V_0}
\left[
(V_0+V_{\rm up})(u-u_c)
+
V_{\rm up}\ln\frac{u-1}{u_c-1}
-
V_0\ln\frac{u}{u_c}
\right],
\]
where $\phi_c$ is fixed by the waterfall condition $m_-^2(\phi_c)=0$. This relation makes explicit how the constant uplift modifies the horizon-crossing field value at fixed $N_*$. At fixed $\phi$, the uplift suppresses $\epsilon_V$ by increasing the total potential energy. At fixed $N_*$, however, this effect is largely compensated by the uplift-induced shift of $\phi_*$. Consequently, the leading attractor behavior remains
\[
n_s
\simeq
1-\frac{2}{N_*},
\qquad
r
\simeq
\frac{12\alpha}{N_*^2}
\left[
1+\delta r(f,u_c,N_*)
\right],
\]
where
\[
f=\frac{V_0}{V_0+V_{\rm up}},
\]
and $\delta r(f,u_c,N_*)$ denotes subleading corrections from the logarithmic terms in the exact $N$--mapping and from the hybrid end point. Thus the dominant control of the tensor amplitude is the curvature parameter $\alpha$, while the uplift produces controlled finite-$N_*$ and end-point corrections rather than an independent leading $f^2$ suppression at fixed $N_*$.

For the benchmark values considered in this work, the predictions remain in the standard red-tilted $\alpha$-attractor regime. For $N_*=60$, one finds $n_s\simeq0.967$--$0.968$, with $r\lesssim4\times10^{-3}$ for $\alpha\le1$. These values are consistent with current CMB constraints, including Planck, ACT DR6, and DESI DR2, but the model should not be interpreted as being designed to reproduce the central ACT-preferred higher-$n_s$ value. Rather, it provides a Planck-compatible red-tilted attractor realization that remains viable in light of the newer data. Depending on $\alpha$, the predicted tensor amplitude may lie within the target range of future $B$-mode searches such as LiteBIRD and CMB-S4~\cite{Hazumi:2019lys,Abazajian:2016yjj}.

The role of the angular component of $T$ is an important part of the effective-field-theory interpretation. In the minimal Kähler realization, ${\rm Im}\,T$ can be lighter than the Hubble scale during inflation. The single-field predictions should therefore be interpreted as applying to the stabilized or trajectory-aligned branch of the model, where the turn rate is negligible and angular fluctuations are not converted into observable isocurvature perturbations. Equivalently, small higher-order Kähler corrections may be used to raise $m_{{\rm Im}\,T}^2/H^2$ above unity without modifying the real-axis plateau at leading order. In the same branch, the expected primordial non-Gaussianity is slow-roll suppressed, with $f_{\rm NL}^{\rm local}=5(1-n_s)/12+\mathcal{O}(\epsilon_i^2)$ and no additional leading bispectrum source from the sequestered uplift or the heavy waterfall fields at CMB horizon exit. If the angular-stabilization or trajectory-alignment assumption is relaxed, the system must be analyzed as a genuine multi-field model, and non-Gaussianity could become model-dependent.

The post-inflationary vacuum structure is controlled by the waterfall transition and by the residual hidden-sector offset. Along the inflationary valley, the constant part of the $U(1)_X$ false-vacuum energy may be absorbed into the effective $V_{\rm up}$. After the waterfall, the charged fields relax to the symmetry-breaking minimum and the positive-definite $U(1)_X$ D-term vanishes. The remaining vacuum energy is therefore of the form
\[
V_{\rm vac}
=
V_{\rm hid}^{\rm res}
+
V_{\rm tune},
\]
which may be tuned to the observed late-time cosmological constant without affecting the inflationary observables, provided the tuning sector remains sequestered from the inflaton during the slow-roll era. The breaking of $U(1)_X$ can lead to cosmic strings, whose tension constrains the waterfall scale. This constraint can be satisfied for sufficiently small $v_D$ or avoided in embeddings where stable strings do not form.

Radiative corrections are under perturbative control. The waterfall-sector masses approach constants on the plateau, while their derivatives with respect to $\phi$ are exponentially suppressed. As a result, the Coleman--Weinberg correction may shift the plateau height, but its contribution to the inflaton slope is exponentially small after the constant part is renormalized. The sequestered Stückelberg sector does not generate an additional inflaton-dependent loop slope in the factorized limit. The one-loop Kähler estimate shows that the mixed $T$--$\Sigma$ correction vanishes in the exactly sequestered branch; if non-sequestered heavy thresholds are present, the induced correction is loop-suppressed and controlled by the threshold parameter $c_{\rm mix}\ln(\Lambda^2/M_{\rm Stk}^2)/(32\pi^2)$. Curved-spacetime corrections to the Coleman--Weinberg potential are also negligible for the heavy thresholds considered here, provided $m_A^2\gg R_4\simeq12H^2$ during the observable stage. After renormalizing the gravitational counterterms, the residual curvature contribution is suppressed by $H^2/(16\pi^2)$ relative to the inflationary energy density and does not destabilize the $\alpha$-attractor plateau.

Several extensions remain worth pursuing. A more microscopic realization of the sequestered uplift, for example in a string-motivated compactification with controlled moduli stabilization, would clarify the origin and stability of the effective parameter $V_{\rm up}$. A detailed study of reheating, including perturbative and nonperturbative decay channels of the waterfall sector into visible degrees of freedom, would sharpen the mapping between $N_*$ and the CMB pivot scale. Finally, the two-field dynamics associated with the angular component of $T$ deserve a dedicated analysis if the stabilizing or trajectory-alignment assumption is relaxed.

In summary, the hybrid $\alpha$-attractor construction with a sequestered constant uplift provides a controlled and analytically tractable supergravity effective framework for inflation. It preserves the characteristic red-tilted attractor prediction for $n_s$, keeps the tensor amplitude governed primarily by $\alpha$, allows a hybrid waterfall exit, and remains compatible with current CMB data while offering a clear target for future polarization measurements.

\section*{Acknowledgments}

The author would like to acknowledge that this work was conducted without external funding. The author appreciates the support of their academic community and peers for their valuable discussions and insights.

\appendix

\section*{Appendix A: Analytic Hubble-flow hierarchy in the slow-roll flow approximation}
\label{app:hubbleflow}

This appendix gives the analytic form of the Hubble-flow hierarchy for the hybrid $\alpha$-attractor potential with a constant uplift, using the slow-roll flow equation employed in Sec.~\ref{sec:background}. The purpose is to provide explicit expressions for $\epsilon_1$, $\epsilon_2$, and $\epsilon_3$ in terms of the auxiliary variable
\[
u\equiv e^{\beta\phi},
\qquad
\beta=\sqrt{\frac{2}{3\alpha}}.
\]
These expressions are useful for evaluating the Stewart--Lyth next-to-leading-order corrections quoted in Sec.~\ref{sec:perturbations}. They should be understood as exact analytic expressions within the first-order slow-roll flow system, not as exact solutions of the full second-order background equation.\footnote{The exact background dynamics would require solving $\ddot\phi+3H\dot\phi+V_{,\phi}=0$ together with $3H^2=\frac12\dot\phi^2+V$. The first-order relation $d\phi/dN\simeq -V_{,\phi}/V$ follows after neglecting $\ddot\phi$ and the kinetic contribution to $H^2$.}

The potential along the stabilized or trajectory-aligned valley is
\begin{align}
V(\phi)
=
U(\phi)+V_{\rm up},
\qquad
U(\phi)
=
V_0
\left(
1-e^{-\beta\phi}
\right)^2 .
\label{eq:app_potential}
\end{align}
Introducing
\begin{align}
u=e^{\beta\phi},
\qquad
\phi=\frac{1}{\beta}\ln u,
\label{eq:app_u_def}
\end{align}
one has
\begin{align}
\frac{d}{d\phi}
=
\beta u\frac{d}{du}.
\label{eq:app_dphi_to_du}
\end{align}
The potential and its first derivative become
\begin{align}
U(u)
=
V_0
\left(
1-\frac{1}{u}
\right)^2
=
V_0
\left(
1-\frac{2}{u}+\frac{1}{u^2}
\right),
\label{eq:app_U_u}
\\
U'(\phi)
=
2\beta V_0
\frac{u-1}{u^2}.
\label{eq:app_Uprime}
\end{align}
The first-order slow-roll flow equation is
\begin{align}
\frac{d\phi}{dN}
\simeq
-\frac{V_{,\phi}}{V}
=
-\frac{U'(\phi)}{U(\phi)+V_{\rm up}},
\label{eq:app_phi_flow}
\end{align}
where $N=\ln a$ is the forward time variable. Therefore
\begin{align}
\frac{du}{dN}
=
\beta u\frac{d\phi}{dN}
=
-\frac{2\beta^2V_0(u-1)}
{uV(u)} ,
\label{eq:app_du_dN_forward}
\end{align}
with
\begin{align}
V(u)
=
V_0
\left(
1-\frac{2}{u}+\frac{1}{u^2}
\right)
+
V_{\rm up}.
\label{eq:app_V_u}
\end{align}
Since the field rolls from larger $u$ to smaller $u$, Eq.~\eqref{eq:app_du_dN_forward} is negative for $u>1$.

The first Hubble-flow parameter in the slow-roll flow approximation is
\begin{align}
\epsilon_1
\simeq
\frac{1}{2}
\left(
\frac{d\phi}{dN}
\right)^2
=
\frac{1}{2}
\left(
\frac{U'(\phi)}{V(\phi)}
\right)^2 .
\label{eq:app_eps1_def}
\end{align}
Using Eq.~\eqref{eq:app_Uprime}, this gives
\begin{align}
\epsilon_1(u)
=
\frac{2\beta^2V_0^2(u-1)^2}
{u^4V(u)^2}.
\label{eq:app_eps1}
\end{align}
The second Hubble-flow parameter is
\begin{align}
\epsilon_2
=
\frac{d\ln\epsilon_1}{dN}
=
\frac{d\ln\epsilon_1}{du}
\frac{du}{dN}.
\label{eq:app_eps2_def}
\end{align}
From Eq.~\eqref{eq:app_eps1},
\begin{align}
\frac{d\ln\epsilon_1}{du}
=
\frac{2}{u-1}
-\frac{4}{u}
-2\frac{V_u}{V},
\qquad
V_u
=
\frac{dV}{du}
=
\frac{2V_0(u-1)}{u^3}.
\label{eq:app_dln_eps1_du}
\end{align}
Therefore
\begin{align}
\epsilon_2(u)
=
-\frac{2\beta^2V_0(u-1)}
{uV(u)}
\left[
\frac{2}{u-1}
-\frac{4}{u}
-\frac{4V_0(u-1)}{u^3V(u)}
\right].
\label{eq:app_eps2_forward}
\end{align}
This expression uses $N=\ln a$ as the forward time variable. If one instead uses $N_{\rm rem}$ as the number of e-folds remaining until the end of inflation, then $d/dN_{\rm rem}=-d/dN$, and the sign of $\epsilon_2$ defined through $d\ln\epsilon_1/dN_{\rm rem}$ is reversed.\footnote{This sign convention is a common source of confusion. The Stewart--Lyth expressions used in Sec.~\ref{sec:perturbations} assume the forward-time convention $N=\ln a$.}

It is useful to display the leading plateau limit. For $u\gg1$,
\begin{align}
V(u)
\simeq
V_0+V_{\rm up},
\qquad
f\equiv\frac{V_0}{V_0+V_{\rm up}},
\label{eq:app_f_def}
\end{align}
and Eq.~\eqref{eq:app_eps1} gives
\begin{align}
\epsilon_1
\simeq
2\beta^2f^2\frac{1}{u^2}.
\label{eq:app_eps1_plateau}
\end{align}
Furthermore,
\begin{align}
\epsilon_2
\simeq
\frac{4\beta^2f}{u}
+
\mathcal{O}\!\left(\frac{1}{u^2}\right).
\label{eq:app_eps2_plateau}
\end{align}
Using the leading relation
\[
u_N\simeq \frac{4fN}{3\alpha},
\]
one obtains
\begin{align}
\epsilon_1
\simeq
\frac{3\alpha}{4N^2},
\qquad
\epsilon_2
\simeq
\frac{2}{N},
\label{eq:app_eps12_N}
\end{align}
which reproduces the standard leading $\alpha$-attractor scaling in the forward-time Hubble-flow convention.

The third Hubble-flow parameter is
\begin{align}
\epsilon_3
=
\frac{d\ln\epsilon_2}{dN}
=
\frac{1}{\epsilon_2}
\frac{d\epsilon_2}{du}
\frac{du}{dN}.
\label{eq:app_eps3_def}
\end{align}
To keep the expression compact, define
\begin{align}
A(u)
=
-\frac{2\beta^2V_0(u-1)}{uV(u)},
\label{eq:app_A_def}
\\
B(u)
=
\frac{2}{u-1}
-\frac{4}{u}
-\frac{4V_0(u-1)}{u^3V(u)}.
\label{eq:app_B_def}
\end{align}
Then
\begin{align}
\epsilon_2(u)
=
A(u)B(u),
\label{eq:app_eps2_AB}
\end{align}
and
\begin{align}
\epsilon_3(u)
=
\left[
\frac{A_u}{A}
+
\frac{B_u}{B}
\right]
\frac{du}{dN}.
\label{eq:app_eps3_AB}
\end{align}
The derivatives entering this expression are
\begin{align}
\frac{A_u}{A}
=
\frac{1}{u-1}
-\frac{1}{u}
-\frac{V_u}{V},
\label{eq:app_A_derivative}
\end{align}
and
\begin{align}
B_u
=
-\frac{2}{(u-1)^2}
+\frac{4}{u^2}
-4V_0
\frac{d}{du}
\left[
\frac{u-1}{u^3V(u)}
\right].
\label{eq:app_B_derivative_start}
\end{align}
The last derivative is
\begin{align}
\frac{d}{du}
\left[
\frac{u-1}{u^3V(u)}
\right]
=
\frac{1}{u^3V(u)}
-\frac{3(u-1)}{u^4V(u)}
-\frac{(u-1)V_u}{u^3V(u)^2}.
\label{eq:app_B_derivative_full}
\end{align}
Equations~\eqref{eq:app_eps3_AB}--\eqref{eq:app_B_derivative_full}, together with Eq.~\eqref{eq:app_du_dN_forward}, give a closed analytic expression for $\epsilon_3(u)$.

In the plateau limit, Eq.~\eqref{eq:app_eps12_N} gives
\begin{align}
\epsilon_3
\simeq
\frac{1}{N}
\label{eq:app_eps3_plateau}
\end{align}
in the forward-time convention used in the Stewart--Lyth expansion. Thus the hierarchy is
\begin{align}
\epsilon_1
\simeq
\frac{3\alpha}{4N^2},
\qquad
\epsilon_2
\simeq
\frac{2}{N},
\qquad
\epsilon_3
\simeq
\frac{1}{N},
\label{eq:app_hierarchy_plateau}
\end{align}
up to finite-$N$ and hybrid-end corrections.

The spectral observables are then obtained from the Stewart--Lyth formulae,
\begin{align}
\mathcal{P}_{\zeta}(k_*)
&=
\frac{H_*^2}{8\pi^2\epsilon_{1*}}
\left[
1-(2\tilde C+1)\epsilon_{1*}
-\tilde C\epsilon_{2*}
\right]
+
\mathcal{O}(\epsilon^2),
\label{eq:app_Pzeta_SL}
\\
n_s
&=
1
-2\epsilon_{1*}
-\epsilon_{2*}
-2\epsilon_{1*}^2
-(2\tilde C+3)\epsilon_{1*}\epsilon_{2*}
-\tilde C\epsilon_{2*}\epsilon_{3*}
+
\mathcal{O}(\epsilon^3),
\label{eq:app_ns_SL}
\\
r
&=
16\epsilon_{1*}
\left[
1+\tilde C(\epsilon_{2*}-2\epsilon_{1*})
\right]
+
\mathcal{O}(\epsilon^3),
\label{eq:app_r_SL}
\\
n_t
&=
-2\epsilon_{1*}
-2(1+\tilde C)\epsilon_{1*}\epsilon_{2*}
+
\mathcal{O}(\epsilon^3),
\label{eq:app_nt_SL}
\end{align}
where
\[
\tilde C=-2+\ln2+\gamma_E\simeq -0.7296.
\]
The pivot value $u_*$ is fixed by the exact slow-roll $N$--mapping derived in Sec.~\ref{sec:background},
\begin{align}
N(u_*;u_c)
=
\frac{1}{2\beta^2V_0}
\left[
(V_0+V_{\rm up})(u_*-u_c)
+
V_{\rm up}\ln\frac{u_*-1}{u_c-1}
-
V_0\ln\frac{u_*}{u_c}
\right].
\label{eq:app_N_mapping}
\end{align}
This procedure ensures that the background trajectory and the perturbation hierarchy are evaluated consistently within the same slow-roll flow approximation.

For numerical work, one may either solve Eq.~\eqref{eq:app_N_mapping} algebraically for $u_*$ at fixed $N_*$ or integrate Eq.~\eqref{eq:app_du_dN_forward} with the end point $u_c$ imposed at $N=0$ in the remaining-$e$-fold convention. In the latter case, care must be taken to transform the sign of the logarithmic derivatives when converting to the forward-time Hubble-flow convention used in Eqs.~\eqref{eq:app_Pzeta_SL}--\eqref{eq:app_nt_SL}.

The compact hierarchy used in the numerical evaluation is therefore
\begin{align}
\epsilon_1(u)
&=
\frac{2\beta^2V_0^2(u-1)^2}
{u^4V(u)^2},
\label{eq:app_compact_eps1}
\\
\epsilon_2(u)
&=
-\frac{2\beta^2V_0(u-1)}
{uV(u)}
\left[
\frac{2}{u-1}
-\frac{4}{u}
-\frac{4V_0(u-1)}{u^3V(u)}
\right],
\label{eq:app_compact_eps2}
\\
\epsilon_3(u)
&=
\left[
\frac{A_u}{A}
+
\frac{B_u}{B}
\right]
\left[
-\frac{2\beta^2V_0(u-1)}
{uV(u)}
\right],
\label{eq:app_compact_eps3}
\end{align}
where $A(u)$ and $B(u)$ are defined in Eqs.~\eqref{eq:app_A_def} and \eqref{eq:app_B_def}. These expressions provide a self-contained analytic implementation of the Hubble-flow hierarchy for the uplifted hybrid $\alpha$-attractor potential within the slow-roll flow approximation.

\bibliography{References}
\bibliographystyle{apsrev4-2}

\end{document}